\documentclass[aip,jcp,10pt,superscriptaddress,
]{revtex4-1}  

\usepackage{bm,bbm}
\usepackage{graphicx}
\usepackage{url}

\usepackage{color}
							
\usepackage[greek,english]{babel}
\usepackage[utf8x]{inputenc}
\usepackage{amsmath}
\usepackage{amssymb,ulem}
\usepackage{graphicx,float,wrapfig}
\usepackage{cancel}
\usepackage{epstopdf}
\usepackage{subfigure} 
\usepackage{color} 
\usepackage{amsthm}

\usepackage{hyperref}
\hypersetup{
    colorlinks = true,
    linkcolor = {black},
    citecolor = {blue},
}

\newcommand{\lat}{\latintext}

\newcommand{\red}{\color{red}}

\begin{document}

\title{\lat \Large{\color{blue} Parameterization of Coarse-grained Molecular Interactions through Potential of Mean Force Calculations and Cluster Expansions Techniques}}


\author{Anastasios Tsourtis}
\email[E-mail: ]{tsourtis@uoc.gr}
\affiliation{Department of Mathematics and Applied Mathematics, University of Crete, Greece}
\author{Vagelis Harmandaris}
\email[E-mail: ]{harman@uoc.gr}
\affiliation{Department of Mathematics and Applied Mathematics, University of Crete, Institute of Applied and Computational Mathematics (IACM), Foundation for Research and Technology Hellas (FORTH), GR-70013, Heraklion, Crete, Greece}
\author{Dimitrios Tsagkarogiannis}
\email[E-mail: ]{D.Tsagkarogiannis@sussex.ac.uk}
\affiliation{Department of Mathematics, University of Sussex, Brighton, BN1 9QH, UK}

\date{\today}

\begin{abstract}
We present a systematic coarse-graining (CG) strategy for many particle molecular systems based on cluster expansion techniques. We construct a hierarchy of coarse-grained Hamiltonians with interaction potentials consisting of two, three and higher body interactions.
The accuracy of the derived cluster expansion based on interatomic potentials is examined over a range of various temperatures and densities and compared to direct computation of pair potential of mean force. 
The comparison of the coarse-grained simulations is done on the basis of the structural properties, against the detailed all-atom data. We give specific examples for methane and ethane molecules in which the coarse-grained variable is the center of mass of the molecule. 
We investigate different temperature and density regimes, and we examine differences between the methane and ethane systems. Results show that the cluster expansion formalism can be used in order to provide accurate effective pair and three-body CG potentials at high $T$ and low $\rho$ regimes. 
In the liquid regime the three-body effective CG potentials give a small improvement, over the typical pair CG ones; however in order to get significantly better results one needs to consider even higher order terms.
\end{abstract}
\maketitle
\section{Introduction}

The theoretical study of complex molecular systems is a very intense research area due to both basic scientific questions and technological applications.~\cite{FrenkelSmitBook}
A main challenge in this field is to provide a direct quantitative link between chemical structure at the molecular level and measurable macroscopic quantities over a broad range of length and time scales. Such knowledge would be especially important for the tailored design of materials with the desired properties, over an enormous range of possible applications in nano-, bio-technology, food science, drug industry, cosmetics etc.

A common characteristic of all complex fluids is that they exhibit multiple length and time scales. 
Therefore, simulation methods across scales are required in order to study such systems. 
On the all-atom level description, classical atomistic models have successfully been used in order to quantitatively predict the properties of molecular systems over a considerable range of length and time scales. \cite{AllenTildesleyBook,FrenkelSmitBook,Harmandaris2003a,TheodorouBook} 
However, due to the broad spectrum of characteristic lengths and times involved in complex molecular systems it is desirable to reduce the required computational cost by describing the system through a small number of degrees of freedom. Thus, coarse-grained (CG) models have been used in order to increase the length and time scales accessible by simulations.\cite{FrenkelSmitBook,IzVoth2005a,tsop1,MulPlat2002,Shell2008,briels,Harmandaris2003a,Harmandaris2006a,Harmandaris2009a,Harmandaris2009b,Johnston2013,IzVoth2005a,Voth2008a,Voth2010,Noid2011,Noid2013,Shell2009,Zabaras2013,DiffusionMapKevrekidis2008,Soper1996,LyubLaa2004}

From a mathematical point of view, coarse-graining is a sub-field of dimensionality reduction;\cite{TheodorouBook} 
there are several statistical methods for the reduction of the degrees of freedom under consideration in a deterministic or stochastic model, such as principal component analysis, polynomial chaos and diffusion maps.\cite{DiffusionMapKevrekidis2008} 
Here we focus our discussion on CG methods based on a combination of recent computational methods and old
theoretical tools from statistical mechanics.     
Such CG models, which are developed by lumping groups of atoms into CG particles and deriving the effective CG interaction potentials directly from more detailed (microscopic) simulations, are capable of predicting \textit{quantitatively} the properties of \textit{specific molecular systems} (see for example refs. \cite{IzVoth2005a,tsop1,MulPlat2002,Shell2008,briels,Harmandaris2009a,Harmandaris2009b,Johnston2013,VagelisReview2014,Voth2010,Noid2013,Shell2009,Zabaras2013,EspanolZuniga2011,BrielsPadding} and references therein).

The most important part in all systematic CG models, based on detailed atomistic data, is to develop rigorous all-atom to CG methodologies that allow, as accurate as possible, estimation of the CG effective interaction. With such approaches the combination of atomistic and hierarchical CG models could allow the study of a very broad range of length and time scales of \textit{specific} molecular systems without adjustable parameters, and by that become truly predictive. \cite{Voth2008a,Harmandaris2009a,Voth2010}
There exists a variety of methods that construct a reduced CG model that approximates the properties of molecular systems based on statistical mechanics. 
For example:

\noindent
(a) In structural, or correlation-based, methods the main goal is to find effective CG potentials that reproduce the {\it pair radial distribution function} $g(r)$, 
and the {\it distribution functions} of bonded degrees of freedom (e.g. bonds, angles, dihedrals) for CG systems with intramolecular interaction potential.~\cite{Soper1996,LyubLaa2004,tsop1,MulPlat2002,Harmandaris2006a,briels}
The CG effective interactions in such methods are obtained using the direct Boltzmann inversion, or reversible work, method~\cite{Nico2014,Harmandaris2006a,Fritz2009,Nico-Review2013} or iterative techniques, such as the iterative Boltzmann inversion, IBI~\cite{Reith2003,MulPlat2002}, and the inverse Monte Carlo, IMC, (or inverse Newton)~\cite{LyubLaa2004,Lyubartsev1995} approach.

\noindent 
(b) Force matching (FM) or multi-scale CG (MSCG) methods \cite{IzVoth2005,IzVoth2005a,Voth2008a,Voth2008b,Noid2011,Voth2013} is a mean least squares problem that considers as observable function  the total force acting on a coarse bead.  

\noindent
(c) The relative entropy (RE) \cite{Shell2008,Shell2009,KP2013}  method employs the minimization of the relative entropy, or Kullback-Leibler divergence, 
between the microscopic Gibbs measure $\mu$ and $\mu^{\theta}$, representing approximations to the exact  coarse space Gibbs measure. 
In this case, the microscopic probability distribution can be thought as the observable. The minimization of the relative entropy is performed through Newton-Raphson approaches and/or stochastic optimization techniques.~\cite{Shell2011,Zabaras2013}

In practice, all above numerical methods are employed to approximate a many body potential of mean force (PMF), $U_{\text{PMF}}$, describing the {\it equilibrium} distribution of CG particles observed in simulations of atomically detailed models. 
Besides the above numerical parametrization schemes, more analytical approaches have also been developed for the approximation of the CG effective interaction, based on traditional liquid state theory and on pair correlation functions.~\cite{Chu2009,Noid2007,Noid2009,Noid2010,Guenza2014,Karniadakis2015,Guenza2013}

Here we discuss an approach for estimating $U_{\text{PMF}}$, and the corresponding effective CG non-bonded potential, based on cluster expansion methods. Such methods originate from the works of Mayer and collaborators \cite{MM} in the 40's.
In the 60's numerous approximate expansions have been further developed \cite{MoHi61, Stell64} for the study of the liquid state.
Later, with the advancement of powerful computational machines, the main focus has been directed on improving the computational methods such as Monte Carlo and molecular dynamics. However, the latter are mostly bulk calculations and they get quite slow for large systems. Reducing the degrees
of freedom by coarse-graining has been a key strategy to
construct more efficient methods, but with many open questions
with respect to error estimation, transferability and
adaptivity of the suggested methods.
Based on recent developments of the mathematical theory of expansion methods in the canonical ensemble \cite{PT12}, our purpose is to combine the two approaches and obtain powerful computational methods, whose error compared to the target atomistic calculations 
can be quantified via rigorous estimates.
In principle, the validity of these methods is limited to the gas regime. Here we examine the accuracy of these methods in different state points. 
This attempt consists of the following: a priori error estimation of the approximate schemes depending on the different regimes, a posteriori error validation of the method from the coarse-grained data and design of related adaptive methods.

In previous years, we have carried out this program for the case of lattice systems, obtaining higher order schemes and a posteriori error estimates \cite{KPRT07}, 
for both short and long range interactions \cite{KPRT14} and designing adaptive methods \cite{KPRT08} and investigating possible strategies for reconstruction of the atomistic information.~\cite{TT10} This is very much in the spirit of the polymer science literature \cite{KM_P, Harmandaris2006a, Harmandaris2009a} and in this paper we get closer by considering off-lattice models. 
The proposed approach is based on typical schemes that are based on isolated molecules.~\cite{Fritz2009,McCoy1998,Nico2014} Here we extend such approaches using cluster expansion tools for deriving CG effective potentials. We start from typical 2-body (pair) effective interaction, but some results can be extended to many-body interactions as well. We also present a detailed theoretical investigation about the effect of higher order terms in obtaining CG effective interaction potentials for realistic molecular systems. 
Then, we show some first results from the implementation of three-body terms on the effective CG potential; a more detailed work on the higher order terms will be given in a forthcoming work.~\cite{THT2017} 

The structure of the paper is as follows: In Section ~\ref{sec:mmodels},
we introduce the atomistic molecular system and its coarse-graining via the definition of the CG map, the n-body distribution function and the corresponding n-body potential of mean force. The cluster expansion based formulation of the
CG effective interaction is presented in Section~\ref{sec:cluster-exp}.
Details about the model systems (methane and ethane) and the simulation considered here are discussed in Section~\ref{sec:simul}.
Results are presented in Section~\ref{sec:results}. 
Finally, we close with Section~\ref{sec:discussion} summarizing the results of this work. 


\section{Molecular Models}\label{sec:mmodels}

\subsection{Atomistic and ``exact'' coarse-grained (CG) description}\label{atcgmodels}

Here we give a short description of the molecular model in the microscopic (all-atom) and mesoscopic (coarse-grained) scale. 
Assume a system of N (classical) atoms (or molecules) in a box 
$\Lambda(\ell):=(-\frac{\ell}{2},\frac{\ell}{2}]^d
\subset\mathbb{R}^d$ (for some $\ell>0$),
at temperature $T$. 
We will also denote the box by $\Lambda$ when we do not need to explicit the dependence on $\ell$.
We consider a configuration $ \mathbf{q}\equiv\{q_1,\ldots,q_N\}$ 
of $N$ atoms, where $q_i$ is the position of the $i^{th}$ atom.
The particles interact via a pair potential
$V:\mathbb R^d\to\mathbb R\cup\{\infty\}$, which is stable and tempered. 
Stability means that there exists a constant
$B\geq0$ such that:
\begin{equation}\label{3}
\sum_{1\leq i<j \leq N} V(q_i-q_j) \geq -BN,
\end{equation}
for all $N$ and all $q_1,...,q_N$. 
Moreover, temperedness requires that
\begin{equation}\label{temper}
C(\beta):= \int_{\mathbb R^d} |e^{-\beta V(r)}-1| dr <\infty.
\end{equation}
where $\beta=\frac{1}{k_{B}T}$ and $k_{B}$ is Boltzmann's constant.
The {\it canonical partition function} of the system is given by
\begin{equation}\label{1a}
Z_{\beta,\Lambda,N}
:=\frac{1}{N!}\int_{\Lambda^N} dq_1\,\ldots dq_N \,e^{-\beta H_{\Lambda}(\mathbf q)},
\end{equation}
where $H_{\Lambda}$ is the Hamiltonian (total energy) of the system confined in a domain $\Lambda$: 
\begin{equation}\label{2}
H_{\Lambda}(\mathbf p, {\mathbf q})
:=\sum_{i=0}^{N}\frac{p^2_i}{2m}+
U({\bf q}).
\end{equation}
By $U(\mathbf{q})$ we denote the total potential energy of the system: 
\begin{equation}\label{25}
U( {\mathbf q}) :=
\sum_{1\leq i<j \leq N} V(q_i-q_j), 
\end{equation}
where for simplicity
we assume periodic boundary conditions on $\Lambda$.
Integrating over the momenta in \eqref{1a}, we get:
\begin{equation}\label{1b}
Z_{\beta,\Lambda,N}
=\frac{\lambda^N}{N!}\int_{\Lambda^N}  dq_1\,\ldots dq_N \,e^{-\beta U(\mathbf q)}=:\lambda^N Z_{\beta,\Lambda,N}^U,
\end{equation}
where $\lambda:=(\frac{2m\pi}{\beta})^{d/2}$.
In the sequel, for simplicity we will consider $\lambda=1$ and identify 
$Z_{\beta,\Lambda,N}\equiv Z_{\beta,\Lambda,N}^U$.
Fixing the positions $q_1$ and $q_2$ of two particles,
we define the two-point correlation function :
\begin{equation}\label{corr_at}
\rho_{N, \Lambda}^{(2), at}(q_1,q_2):=\frac{1}{(N-2)!}\int dq_3 \ldots dq_N 
\frac{1}{Z_{\beta,\Lambda,N}}
e^{-\beta U(\mathbf q)}.
\end{equation}
It is easy to see that in the thermodynamic limit the leading order
is $\rho^2$, where $\rho=\frac{N}{|\Lambda|}$
and $|\Lambda|$ is the volume of the box $\Lambda$.
Thus, it is common to define the following order one quantity
$g(r):=\frac{1}{\rho^{2}}\rho_{N, \Lambda}^{(2), at}(q_1,q_2)$, for
$r=|q_1-q_2|$.
More generally, for $n\leq N$, we define the  $n$-body 
version
\begin{equation}\label{eq:gofr_atom}
g^{(n)}(q_1,\dots,q_n) = \frac{1}{(N-n)! \rho^n}\int_{\Lambda^{N-n}} dq_{n+1}\dots dq_N\!
\frac{1}{Z_{\beta,\Lambda,N}}e^{-\beta U(\mathbf q)},
\end{equation}
and from that the order $n$
potential of mean force (PMF), $U_{\text{PMF}}(q_1,\dots,q_n)\!$, \cite{Kirkwood1935, McQ} given by
\begin{equation}
U_{\text{PMF}}(q_1,\dots,q_n) := -\frac{1}{\beta} \log g^{(n)}(q_1,\dots,q_n) .
\end{equation}

We define the coarse-graining map $T:(\mathbb R^d)^N\to (\mathbb R^d)^M$
on the microscopic state space, 
given by $T: \mathbf q\mapsto T(\mathbf q)\equiv (T_1(\mathbf 	q),\ldots, T_M(\mathbf q))\in\mathbb R^M$, which
determines the $M$ $(M<N)$ CG degrees of freedom as a function  of the atomic configuration $\bf q$. 
We call ``CG particles'' the elements of the coarse space  with positions $\mathbf{r}\equiv\{r_1,\ldots,r_M\}$. 
The effective CG potential energy is defined by
\begin{equation}\label{cg}
U_{\text{eff}}(r_1,\ldots,r_M):=-\frac{1}{\beta}\log\int_{\{T\mathbf{q}=\mathbf{r}\}}
dq_1\,\ldots dq_N \,e^{-\beta U(\mathbf q)},
\end{equation}
where the integral is over all atomistic configurations that correspond to a specific CG one using the coarse-graining map.
In the example we will deal with later, the configuration $\mathbf r$ will represent the centers of mass of groups
of atomistic particles.
This coarse graining gives rise to a series of 
multi-body effective potentials of one, two, up to $M$-body interactions, which are unknown functions of the CG configuration. Note also that by the construction of the CG potential in \eqref{cg} the partition function is the same:
\begin{equation}\label{part}
Z_{\beta,\Lambda,N}=
\int dr_1\ldots dr_M
\int_{\{T\bf q=\bf r\}}
d\mathbf q 
e^{-\beta U(\mathbf q)}=
\int dr_1\ldots dr_M
e^{-\beta U_{\text{eff}}(r_1,\ldots,r_M)}
=:Z^{cg}_{\beta,\Lambda,M}
\end{equation}

The main purpose of this article is to give a systematic way
(via the cluster expansion method) of constructing controlled approximations of $U_{\text{eff}}$ that can be efficiently computed and at the same time we have a quantification of the corresponding error for both {\it ``structural"} and {\it ``thermodynamic"} quantities.
By structural we refer to $g(r)$, while by thermodynamic to the pressure and the free energy. Note that both depend on the partition function, but they can also be related \cite{McQ} to each other as follows: 
\begin{equation}\label{press}
\beta p=\rho-\frac{\beta}{6}\rho^2 \int_0^{\infty}r u'(r)g(r)4\pi r^2 dr,
\end{equation}
at least for the case of pair-interaction potentials.

\subsection{Coarse-grained approximations}\label{cgapp}

As mentioned above there are several methods in the literature that give approximations to the effective (CG) interaction
potential $U_{\text{eff}}$ as defined in \eqref{cg}.
Below we list some of them without claim of being exhaustive:

\noindent
(a) The `correlation-based (eg. DBI, IBI and IMC) methods that use the {\it pair radial distribution function} $g(r)$, related to the two-body potential of mean force for the intermolecular interaction potential, as well as distribution functions of bonded degrees of freedom (e.g. bonds, angles, dihedrals) for CG systems with intramolecular interaction potential.\cite{Soper1996,LyubLaa2004,tsop1,MulPlat2002,Harmandaris2006a,briels}
These methods will be further discussed below.

\noindent 
(b) Force matching (FM) methods \cite{IzVoth2005,IzVoth2005a,Noid2011} in which the observable function is the average force acting on a CG particle. 
The CG potential is then determined from  {\it  atomistic force  information} through a least-square minimization principle, to variationally project the force corresponding to the potential of mean force onto a force that is defined by the form of the approximate potential.

\noindent
(c) Relative entropy (RE)\cite{Shell2008,Shell2009,Zabaras2013} type methods that produce optimal CG potential parameters by minimizing the relative entropy, Kullback-Leibler divergence 
between the atomistic and the CG {\it Gibbs measures} sampled by the atomistic model.

In addition to the above numerical methods, analytical works for the estimation of the effective CG interaction, based on integral equation theory, have also been developed ~\cite{Guenza}.
A  brief review and categorization of parametrization  methods at equilibrium is given in references~\cite{Noid2013,KCTKPH2016}.

The correlation-based iterative (e.g. IBI and IMC) methods use the fact that for a pair interaction $u(r)$, by plugging the virial expansion of $p$ in powers of $\rho$ into \eqref{press} and
comparing the orders of $\rho$, one obtains that
\begin{equation}\label{relation}
g(r)=e^{-\beta u(r)}\gamma(r),\quad \gamma(r)=1+c_1(r) \rho+ c_2(r) \rho^2+\ldots
\end{equation}
Given the atomistic {\it ``target"} $g(r)$ from a free (i.e., without constraints) atomistic run, 
by inverting \eqref{relation} and neglecting the higher order terms of $\gamma(r)$ one can
obtain a first candidate for a pair coarse-grained potential $u(r)$. Then, one calculates
the $g(r)$ that corresponds to the first candidate and by iterating this procedure
eventually obtains the desired two-body coarse-grained potential.
This iteration should in principle converge since there exists
a pair interaction that can be reconstructed from a given correlation function \cite{}.
However, this is only an approximation (accounting for
the neglected terms of order $\rho$ and higher in the expansion of $\gamma(r)$) since we know that
the ``true" CG interaction potential should be multi-body,
as a result of integrating atomistic degrees of freedom.
Hence, having agreement on $g(r)$ does
not secure proper thermodynamic behaviour and several methods have been
employed towards this direction, see for example refs~\cite{MulPlat2002,Louis02,Guenza2014} and the references within.

In order to maintain the correct thermodynamic properties, our approach in this paper is based on cluster expanding \eqref{cg} with respect to some small but finite parameter $\epsilon$
depending on the regime we are interested in. 
For technical reasons we will focus on low density - high temperature regime.
As it will be explained in detail in the next section, the resulting cluster expansion provides us with
a hierarchy of terms:
\begin{equation}\label{approx}
U_{\text{eff}}=U^{(2)}+U^{(3)}+O(\epsilon^3),
\qquad
U^{(2)}(r_1,\ldots, r_M):=\sum_{i,j}W^{(2)}(r_i,r_j),
\quad
U^{(3)}(r_1,\ldots, r_M):=
\sum_{i,j,k}W^{(3)}(r_i,r_j,r_k),
\quad
\text{etc},
\end{equation}
together with the corresponding error estimates.

The above terms can in principle be calculated independently via fast atomistic simulations of 2, 3, etc. molecules, in the spirit of the conditional reversible work CRW method. 
In more detail, the effective non-bonded (two-body) CG potential can be computed as follows:

\noindent
(a) One method is by fixing the distance $r_{1,2}:=r_1-r_2$ between two molecules and perform molecular dynamics with such forces that maintain the fixed distance $r_{1,2}$.
In this way we sample over the constrained phase space and obtain the conditioned 
partition function as in \eqref{cg}.
Then, by integration of the constrained force the two-body effective potential can be obtained.

\noindent
(b) Alternatively, by inverting $g(r)$ in \eqref{relation} for two isolated molecules, the two-body effective potential can be directly obtained, since for such a system $\gamma(r)=1$. 

Here we examine both methods, see Figure~\ref{fig:fig3}. 
Note also that the validity of cluster expansion provides rigorous expansions
for $g(r)$, the pressure and the other relevant quantities.
Hence, with this approach we can have {\it a priori} estimation of the errors made in \eqref{relation}.
Another benefit of the cluster expansion is that the error terms can  be written in terms of the coarse-grained quantities allowing for a posteriori error estimates and the design of adaptive methods \cite{KPRT08}; see also discussion in Section \ref{sec:discussion}.

\section{Cluster expansion}\label{sec:cluster-exp}
The cluster expansion method originates from the work of Mayer and collaborators, see ref.~\cite{MM}
for an early review, and consists of expanding the logarithm of the partition function in an absolutely convergent series of an appropriately chosen small but finite parameter. 
Here we will adapt this method to obtain an expansion of the conditioned partition function \eqref{cg}.

For the purpose of this article we assume that the CG map $T$ is a product
$T=\otimes_{i=1}^M T^i$ creating $M$ groups of $l_1, \ldots, l_M$ particles each. 
We index the particles in the $i^{th}$ group of the coarse-grained variable $r_i$
by $k^i_1, \ldots, k^i_{l_i}$.
We also denote them by $\mathbf{q}^i:=(q_{k^i_1}, \ldots, q_{k^i_{l_i}})$,
for $i=1,\ldots,M$.
Then \eqref{cg} can be written as:
\begin{equation}\label{cg2}
U_{\text{eff}}(r_1,\ldots,r_M):=-\frac{1}{\beta}\log\prod_{i=1}^M \lambda^i(\{T^i\mathbf{q}^i=r_i\})
-\frac{1}{\beta}\log\int \prod_{i=1}^M 
\mu(d\mathbf{q}^i;r_i)
e^{-\beta U(\mathbf q)},
\end{equation}
where, for simplicity, we have introduced the normalized conditional measure:
\begin{equation}\label{mu}
\mu(d\mathbf{q}^i;r_i):=\frac{1}{l_i!}
dq_{k^i_1}\ldots dq_{k^i_{l_i}}\frac{\mathbf 1_{\{T^i\mathbf{q}^i=r_i\}}}{\lambda^i(\{T^i\mathbf{q}^i=r_i\})},
\end{equation}
and by $\lambda^i$ we denote the measure $\frac{1}{l_i!}dq_{k^i_1}\ldots dq_{k^i_{l_i}}$.
To perform a cluster expansion in the second term of \eqref{cg2} 
we rewrite the interaction potential as follows:
\begin{equation}\label{hamil}
U(\mathbf q)=
\sum_{i<j} \bar V(\mathbf{q}^i,\mathbf{q}^j),
\quad\text{where}\quad
\bar V(\mathbf{q}^i,\mathbf{q}^j):=\sum_{m=1}^{l_i}\sum_{m'=1}^{l_{j}} V(|q_{k^i_m}-q_{k^j_{m'}}|).
\end{equation}
Then, we have
\begin{eqnarray}\label{CE}
e^{-\beta U(\mathbf q)} & = &
\prod_{i<j} \left(
1+ e^{-\beta \bar V(\mathbf{q}^i,\mathbf{q}^j)}-1
\right) \nonumber\\
&=&
\sum_{\substack{V_1,\ldots,V_m \\ |V_i|\geq 2, V_i\subset\{1,\ldots,N\}}}
\prod_{l=1}^m\sum_{g\in\mathcal C_{V_l}}\prod_{\{i,j\}\in E(g)} f_{i,j}(\mathbf{q}^i,\mathbf{q}^j),
\quad
\text{where}
\quad
f_{i,j}(\mathbf{q}^i,\mathbf{q}^j)
:=e^{-\beta \bar V(\mathbf{q}^i,\mathbf{q}^j)}-1,
\end{eqnarray}
where for $V\subset\{1,\ldots,N\}$, we denote by $\mathcal C_{V}$
the set of {\it connected} graphs on the set of vertices with labels in $V$.
Furthermore, for $g\in \mathcal C_{V}$, we denote by $E(g)$ the set of its edges.
\begin{figure}
	\resizebox{0.5\columnwidth}{!}
	{\includegraphics{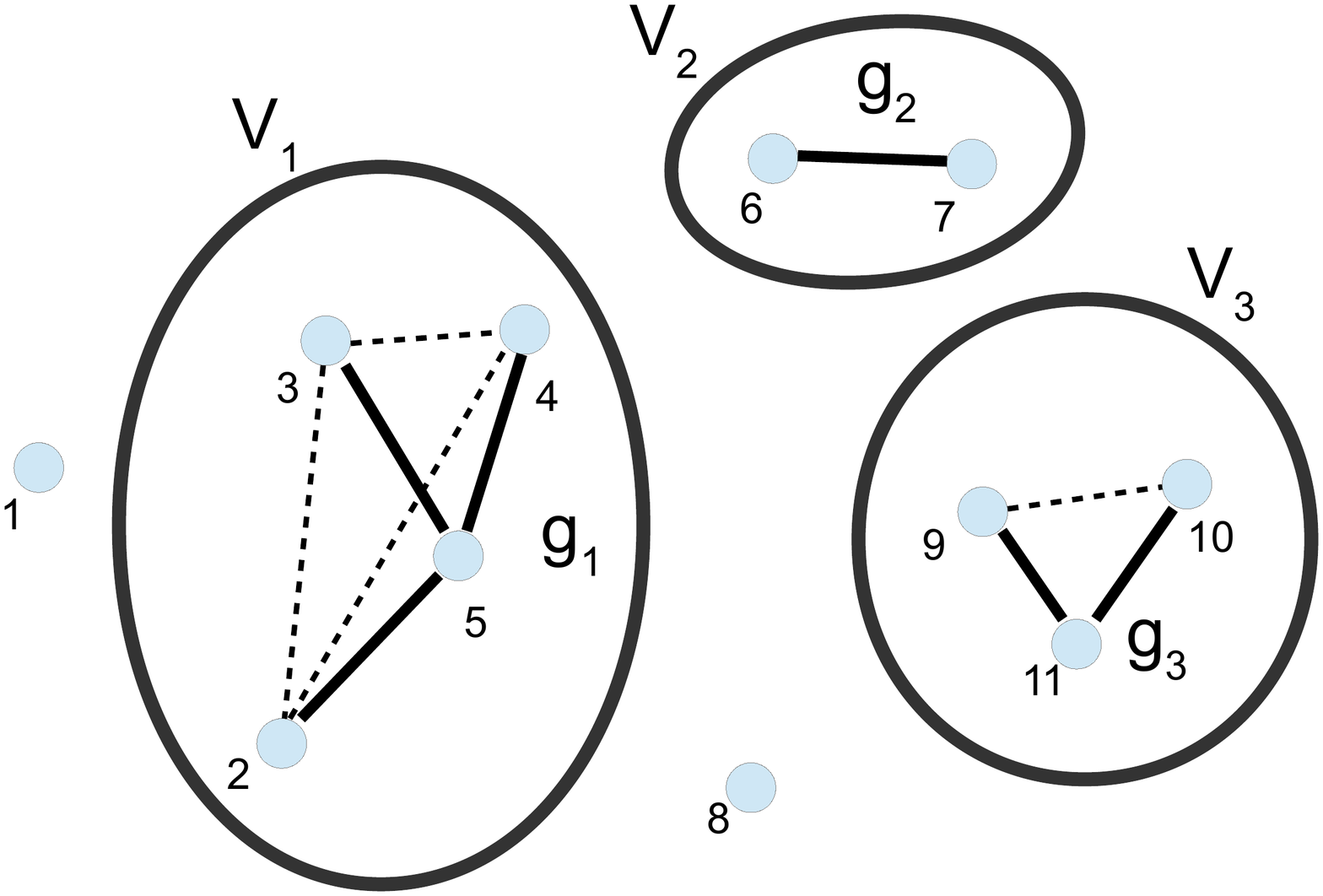}}
	\caption{Visualization of the partition in \eqref{CE} for non-intersecting sets $V_1=\{2,3,4,5\}$, $V_2=\{6,7\}$, $V_3=\{9,10,11\}$ in each of which we display by solid lines the connected graphs $g_i\in\mathcal C_{V_i}$, $i=1,2,3$.}
	\label{fig:fig2}
\end{figure} 
Since $\mu$ in \eqref{mu} is a normalized measure, from \eqref{cg2} we obtain:
\begin{eqnarray}\label{cg3}
U_{\text{eff}}(r_1,\ldots,r_M) & = &
-\frac{1}{\beta}\log\prod_{i=1}^M \lambda^i(\{T^i\mathbf{q}^i=r_i\})
-\frac{1}{\beta}\log
\sum_{\substack{V_1,\ldots,V_m \\ |V_i|\geq 2, V_i\subset\{1,\ldots,N\}}}
\prod_{l=1}^m
\zeta(V_i)\nonumber\\
&=&
-\frac{1}{\beta}\log\prod_{i=1}^M \lambda^i(\{T^i\mathbf{q}^i=r_i\})
-\frac{1}{\beta}
\sum_{V\subset\{1,\ldots,N\}}\zeta(V)+\frac{1}{\beta}\sum_{\substack{V,V':\\ V\cap V'=\varnothing}}\zeta(V)\zeta(V')+\ldots,
\end{eqnarray}
where $\zeta(V):=\int \sum_{g\in\mathcal C_{V}}\prod_{\{i,j\}\in E(g)} f_{i,j}(\mathbf{q}^i,\mathbf{q}^j) d\mathbf q_V$ with $\mathbf q_V:=\{\mathbf q^i\}_{i\in V}$, is a function over the atomistic details of the system. 
Note that the above expression involves a sum over all possible pairs, triplets etc. which is a convergent series for values of the density $\rho=\frac{N}{|\Lambda|}$ and of the inverse temperature $\beta$ such that
$\rho C(\beta)<c_0$, where $C(\beta)$ is defined in \eqref{temper} and 
$c_0$ is a known small positive constant.\cite{PT12}
If we simplify the sum in \eqref{cg3}
one can obtain \cite{PT12} expansion \eqref{approx} where
\begin{equation}\label{two-body}
W^{(2)}(r_1,r_2):=-\frac{1}{\beta}\int \mu(d{\bf q}^1;r_1)\, \mu(d{\bf q}^2;r_2) \,
f_{1,2}(\mathbf{q}^1,\mathbf{q}^2)
\end{equation}
and
\begin{equation}\label{three-body}
W^{(3)}(r_1,r_2,r_3):=-\frac{1}{\beta}\int \mu(d{\bf q}^1;r_1)\,\mu(d{\bf q}^2;r_2)\,\mu(d{\bf q}^3;r_3)\,
f_{1,2}(\mathbf{q}^1,\mathbf{q}^2)\,
f_{2,3}(\mathbf{q}^2,\mathbf{q}^3)\,
f_{3,1}(\mathbf{q}^3,\mathbf{q}^1).
\end{equation}
Recall also the definition of $f_{i,j}$ in \eqref{CE}.



\subsection{Full calculation of the PMF}

Notice that the potentials $W^{(2)}$ and $W^{(3)}$ in \eqref{two-body}
and \eqref{three-body}, respectively, have been expressed via the Mayer functions $f_{i,j}$.
However, the full effective interaction potential between two CG particles  can be directly defined as the (conditional) two-body PMF given by  
\begin{equation}\label{MC2}
W^{(2), \text{full}}(r_1,r_2):=-\frac{1}{\beta}\log\int \mu(d{\bf q}^1;r_1)\, \mu(d{\bf q}^2;r_2) \,
e^{-\beta \bar V({\bf q}^1,{\bf q}^2)}.
\end{equation}
By adding and subtracting $1$, we can relate it to 
\eqref{two-body}:
\begin{eqnarray}\label{alt}
-\beta W^{(2), \text{full}}(r_1,r_2) & = &
\log\int \mu(d{\bf q}^1;r_1)\, \mu(d{\bf q}^2;r_2) \,
e^{-\beta \bar V({\bf q}^1,{\bf q}^2)} =
\log(1+\int \mu(d{\bf q}^1;r_1)\, \mu(d{\bf q}^2;r_2) \,
f_{1,2}(\mathbf{q}^1,\mathbf{q}^2))\nonumber\\
& = &
\int \mu(d{\bf q}^1;r_1)\, \mu(d{\bf q}^2;r_2) \,
f_{1,2}(\mathbf{q}^1,\mathbf{q}^2)
-\frac 12
\left(
\int \mu(d{\bf q}^1;r_1)\, \mu(d{\bf q}^2;r_2) \,
f_{1,2}(\mathbf{q}^1,\mathbf{q}^2)
\right)^{2}
+\ldots
\end{eqnarray}
Higher order terms in the above equation are expected to be less/more important in high/low temperature.

Similarly, for three CG degrees of freedom $r_1, r_2, r_3$, the full PMF is
given by 
\begin{equation}\label{MC3}
W^{(3), \text{full}}(r_1,r_2,r_3):=-\frac{1}{\beta}
\log\int \mu(d{\bf q}^1;r_1)\, \mu(d{\bf q}^2;r_2) \mu(d{\bf q}^3;r_3)\,
e^{-\beta \sum_{1\leq i < j \leq 3}\bar V({\bf q}^i,{\bf q}^j)}.
\end{equation}
By adding and subtracting $1$ we can relate it to 
\eqref{two-body} and \eqref{three-body}
(in the following we simplify notation by not explicitly showing   the dependence on the atomistic configuration and neglecting the normalized conditional measure):
\begin{eqnarray}
e^{-\beta W^{(3),\text{full}}} & = & \int e^{-(V_{12}+V_{13}+V_{23})}\nonumber\\
& = & 1+ \int f_{12}+\int f_{13}+\int f_{23}+\int f_{12}f_{13}+\int f_{13}f_{23}+\int f_{12}f_{23}+\int f_{12}f_{23} f_{13},
\end{eqnarray}
which implies that
\begin{eqnarray}
W^{(3),\text{full}} & = & -\frac{1}{\beta}\left[
\int f_{12}+\int f_{13}+\int f_{23}+\int f_{12}f_{23} f_{13}+\right.\nonumber\\
&& \left.
\int f_{12}f_{13}+\int f_{13}f_{23}+\int f_{12}f_{23}-(\int f_{12} \int f_{13}+\int f_{13}\int f_{23}+\int f_{12}\int f_{23})
\right]+\ldots
\end{eqnarray}

In principle, we can rewrite \eqref{approx} with respect to $W^{(2),\text{full}}$ and $W^{(3),\text{full}}$.
Note however, that both of these terms contain the coarse-grained two-body interactions, hence in order to avoid double-counting, when we use both, we have to appropriately subtract the two-body contributions. For some related results, see also the discussion about Figure~\ref{fig:fig11}.

\subsection{Thermodynamic consistency}

As already mentioned, several coarse-graining
strategies lack of thermodynamic consistency, see also
the discussion by Louis \cite{Louis01} and Guenza \cite{Guenza2014}.
On the other hand, by construction, the
cluster expansion approach gives
quantified approximations to the correct
thermodynamic behaviour.
Hence, from \eqref{approx}, by considering only the
two-body contribution,
for the finite volume free energy we have that
\begin{equation}\label{free_energy}
-\frac{1}{\beta |\Lambda|}\log Z_{\beta,\Lambda,N} = 
-\frac{1}{\beta |\Lambda|}\log\int dr_1\ldots dr_M
e^{-\beta U^{(2)}}+\frac{1}{\beta |\Lambda|} O(\epsilon^3)),
\end{equation}
where the error is uniform in $N$ and $|\Lambda|$ and
negligible in the limit.
Thus, the approximation $U^{(2)}$ of the CG Hamiltonian
implies a good approximation of the free energy.
Similarly,  for the {\it pressure} as a function of
the activity $z$, we have:
\begin{equation}\label{pressure}
\frac{1}{\beta |\Lambda|}\log\sum_{N\geq 0}z^N Z_{\beta,\Lambda,N}
=\frac{1}{\beta |\Lambda|}\log\sum_{N\geq 0}z^N
\int dr_1\ldots dr_M
e^{-\beta U^{(2)}}
+\frac{1}{\beta |\Lambda|}
O(\epsilon^3).
\end{equation}
Both quantities have limits given by absolutely convergent series with 
respect to $\rho=N/|\Lambda|$ for the first and $z$ or $\rho$ for the second.
As a side remark, let us mention that
in order to compute them we have two options:
the first is to use \eqref{free_energy} and calculate the integral
$\int dr_1\ldots dr_M e^{-\beta U^{(2)}}$ using molecular dynamics.
Alternatively, we can use the corresponding expansions
-
e.g. for the free energy we would obtain
\cite{PT14}
\begin{equation}\label{virial}
-\frac{1}{\beta |\Lambda|}\log Z_{\beta,\Lambda,N} 
=
\rho(\log\rho-1)+\sum_{n\geq 1} \beta_{\Lambda}\rho^n+\text{finite volume errors}
\end{equation}
- and compute the coefficients $\beta_{\Lambda}$.
The latter are not bulk computations as they involve $2$, $3$, etc
particles so they are rather efficient, at least up to some order.

\subsection{Pair correlation function}

Recalling the coarse-grained map $T$ from the previous section, we fix two centers of mass $r_1$ and $r_2$
and integrate over all atomistic configurations so that
the first two groups $\mathbf q^1$ and $\mathbf q^2$
of atomistic configurations
have the above fixed centers of mass.
Partitioning the $N$ particles into $M$ groups of $l_1,\ldots,l_M$ particles
and choosing two of them (indexed by $1$ and $2$) to be the fixed ones,
we define the ``projected'' correlation function
at the coarse-grained scale as follows:
\begin{eqnarray}\label{pair_proj_cg}
\rho_{N,\Lambda}^{(2), proj}(r_1, r_2) & := & 
\int_{\{T_1({\mathbf q}^1)=r_1, \, T_2({\bf q}^2)=r_2\}}
\prod_{i=1}^M \lambda^i(d\mathbf q^i) \, 
\frac{1}{Z_{\beta,\Lambda,N}}
e^{-\beta U(\mathbf q)}\nonumber\\
&=& \int dr_3 \ldots dr_M 
\int \prod_{i=1}^M 
\mu(d\mathbf{q}^i;r_i)
\frac{1}{Z_{\beta,\Lambda,N}}
e^{-\beta U(\mathbf q)}\nonumber\\
& = &
\int dr_3 \ldots dr_M
\frac{1}{Z^{cg}_{\beta,\Lambda,M}} 
e^{-\beta U_{\text{eff}}(r_1,\ldots,r_M)}.
\end{eqnarray}
Hence, using \eqref{approx} we can construct coarse-grained 
approximations for the correlation functions as well.
Alternatively,
as a corollary of the cluster expansion, we can 
write \eqref{corr_at} as
a convergent power series with respect to the density.
These are old results \cite{Stell64} for which the
convergence has also been proved recently in the context of the canonical ensemble.~\cite{KT}
In the limit $N\to \infty$, $\Lambda\to \mathbb R^d$ such that $\frac{N}{|\Lambda|}=\rho$,
we obtain:
\begin{equation}\label{gofr}
g(r)=
e^{-\beta \bar V({\bf q}^1- {\bf q}^2)}
\left[
1+\rho C_3({\bf q}^1,{\bf q}^2)
+\rho^2 C_4({\bf q}^1,{\bf q}^2)
+\ldots
\right],
\qquad r:=T({\bf q}^1)-T({\bf q}^2),
\end{equation}
where
\begin{equation}\label{C3}
C_3({\bf q}^1,{\bf q}^2):= \int_{\Lambda}d{\bf q}_3 \, f_{1,3} f_{3,2},
\qquad 
f_{i,j}:=e^{-\beta \bar V({\bf q}_i-{\bf q}_j)}-1
\end{equation}
and
\begin{equation}\label{C4}
C_4({\bf q}^1,{\bf q}^2):  =  \int dq_3\, dq_4 \, f_{1,3}  f_{3,4} f_{4,2}
+ 4 \int dq_3\, dq_4 \, f_{1,3} f_{3,4} f_{1,4} f_{4,2}
 + \int dq_3 \, dq_4 f_{1,3} f_{3,2} f_{1,4} f_{4,2}
+\int dq_3 \, dq_4 \, f_{1,3} f_{1,4} f_{2,3} f_{2,4} f_{3,4}
\end{equation}
Note that this formula could also be used at
the coarse-grained level with the pair coarse-grained potential
$W^{(2)}$, giving an alternative way to
compute it.

\section{Model and Simulations}\label{sec:simul}

\subsection{The model}
A main goal of this work, as mentioned before, is to examine the parameterization of a coarse-grained model using the cluster expansion formalism described above for simple realistic molecular systems; in this work we study liquid methane and ethane. 
In more detail, we consider $N$ molecules of $CH_4$ and we denote as $\bar{\bf q}\equiv\{\bar q_1,\ldots,\bar q_N\}$ to be the positions of the $N$ many carbons and ${\bf q}_i\equiv\{q_{i,1},\ldots, q_{i,4}\}$ be the positions of the $4$ hydrogens that
correspond to the $i^{th}$ carbon.
We have two types of interactions, namely the {\it bonded} with (many body) interaction potential $V_b$ and the {\it non-bonded} with pair interaction potential $V_{nb}$. The latter are of Lennard-Jones type between all possibilities: $C-C$, $C-H$ and $H-H$ (with different coefficients), i.e., $V_{nb}=V_{CC}+V_{CH}+V_{HH}$.
In the model used here the non-bonded interactions within the same $CH_4$ molecule are excluded. 

The microscopic canonical partition function is given by
\begin{equation}\label{100}
Z_{CH_4}=\frac{1}{N!}\int_{\Lambda^N} d\bar{\bf q}\,(\frac{1}{4!})^N
\int_{\Lambda^{4N}}\prod_{i=1}^N d{\bf q}_i
e^{-\beta\left(\sum_{i=1}^N V_b(\bar q_i,{\bf q}_i)+U_{nb}(\bar{\bf q},{\bf q}_1,\ldots,{\bf q}_N)\right)},
\end{equation}
where $U_{nb}$ is a pair potential of all possible pairs among 
$\bar{\bf q},{\bf q}_1,\ldots,{\bf q}_N$, all of L-J type (eventually with different parameters).
Note also that since only the $4$ particles of $H$ are indistinguishable, we
have introduced the factor $1/4!$ for each molecule.

\medskip
We are interested in computing the effective Hamiltonian when only the
centers of mass of the $N$ many molecules are prescribed.
Hence, let us introduce a map $T:\Lambda^5\to\Lambda$ 
which gives the center of mass of a molecule consisting of an atom of $C$ 
together with the prescribed $4$ atoms of $H$ which are linked to $C$ by the bonded interactions, i.e., by denoting ${\bf\bar q}_i\equiv (\bar q_i,{\bf q}_i)$ we have:
\begin{equation}\label{101}
T({\bf \bar q}_i):=\frac{1}{m_C+4 m_H}(m_C \bar q_i+ m_H\sum_{j=1}^4 q_{i,j}).
\end{equation}
We introduce the variables $r_1,\ldots, r_N$ for the centers of mass.
Our goal is to find the effective potential $U_{\text{eff}}(r_1,\ldots,r_N)$.
We define the ``bonded" (normalized) prior measure by
\begin{equation}\label{103}
d\hat\mu_b({\bf \bar q}_i;r_i):=\frac{1}{Z_b(r_i)}d{\bf \bar q}_i
\mathbf 1_{T({\bf \bar q}_i)=r_i}e^{-\beta V_b({\bf \bar q}_i)},\quad
Z_b(r_i):=\frac{1}{4!}\int_{\Lambda^5}d{\bf \bar q}_i
\mathbf 1_{T({\bf \bar q}_i)=r_i}
e^{-\beta V_b({\bf \bar q}_i)}.
\end{equation}
Note that here we could have also included possible non-bonded interactions between atoms of the same molecule. This would be important for the case of coarse-graining a molecule with intra-molecular non-bonded interactions; for the methane molecule studied here such interactions do not exist.
Then, from \eqref{100} we obtain:
\begin{equation}\label{102}
Z_{CH_4} = \frac{1}{N!}\int_{\Lambda^N}dr_1\ldots dr_N\,
\prod_{i=1}^N Z_b(r_i)
\int \prod_{i=1}^N d\hat\mu_b({\bf \bar q}_i;r_i)
e^{-\beta U_{nb}({\bf \bar q}_1,\ldots,{\bf \bar q}_N)}.
\end{equation}
The effective free energy is defined by:
\begin{equation}\label{103.0}
e^{-\beta U_{\text{eff}}(r_1,\ldots,r_M)}:=
\prod_{i=1}^N Z_b(r_i)
\int \prod_{i=1}^N d\hat\mu_b({\bf \bar q}_i;r_i)
e^{-\beta U_{nb}({\bf \bar q}_1,\ldots,{\bf \bar q}_N)},
\end{equation}
for which we can construct approximations
following formula \eqref{cg3}.
A similar analysis holds for ethane as well.

The total (atomistic) potential energy $V({q})$, for both methane and ethane, is defined by
\begin{equation}
V({q}) = V_{bond}({q}) + V_{angle}({q}) + V_{LJ}({q}) \ .
\end{equation}
where $V_{bond}({q}), V_{angle}({q})$ are quadratic intramolecular potential functions of the bonds and angles respectively. $V_{LJ}({q})$ is the non-bonded potential as defined in the previous subsection.
The parameters values of $CH_4$ are summarized in Table \ref{tab:meth_params}.
\begin{table}[!hbt]
	\centering
	\begin{tabular}{ | l || c | r | r | }
		\hline
		& $\epsilon_{LJ} \tiny{[\frac{Kcal}{mol} ]}$ & $\sigma_{LJ}$ \tiny{[\AA]} & $r_{cut}$ \tiny{[\AA]}\\
		\hline
		$C-C$ & 0.0951 & 3.473 & 15.0  \\
		$C-H$ & 0.0380 & 3.159 & 15.0  \\
		$H-H$ & 0.0152 & 2.846 & 15.0\\
		\hline
	\end{tabular}
	\begin{tabular}{ |c|c|c|c|}
		\hline
		$K_{b}$ \tiny{[$\frac{Kcal}{mol \AA^{2}}$]} & $r_0$ \tiny{[$\AA$]}  & $K_{\theta}$ \tiny{[$\frac{Kcal}{mol \cdot deg^{2}}$]} & $\theta_0$ \tiny{[rad]} \\
		\hline
		700 & 1.1 & 100  & 1.909  \\
		\hline
	\end{tabular}
	\caption{Non-bonded $LJ$ coefficients as well as bond and angle coefficients for methane. \cite{Dreiding1990}}
	\label{tab:meth_params}
\end{table}

The more simple, non-spherically symmetric ethane molecule consists of one rigid bond connecting two united atom $CH_{3}$ beads. Table \ref{tab:ETH_params} summarizes this model.
\begin{table}[!hbt]
	\centering
	\begin{tabular}{ | l || c | r | r | }
		\hline
		 & $\epsilon_{LJ} \tiny{[\frac{Kcal}{mol} ]}$ & $\sigma_{LJ}$ \tiny{[\AA]} & $r_{cut}$ \tiny{[\AA]}\\
		\hline
		 $CH_{3}-CH_{3}$& 0.194726 & 3.75 & 14.0  \\
		\hline
	\end{tabular}
\caption{Non-bonded $LJ$ coefficients for ethane. \cite{Wick_Siepmann_Trappe_ETH}}
	\label{tab:ETH_params}
\end{table}

\subsection{Simulations}

The simplest system to simulate is the one with only two interacting methane, or ethane, molecules in vacuum. This is a reference system for which the many-body PMF is equal to the two-body one. In addition we have also simulated the corresponding liquid systems. The atomistic and CG model methane systems were studied through molecular dynamics and Langevin dynamics (LD) simulations. All simulations were conducted in the NVT ensemble. For the MD simulations the Nose-Hoover thermostat was used. 
Langevin dynamics models a Hamiltonian system which is coupled with a thermostat. \cite{lelievre2010} The thermostat serves as a reservoir of energy.
The densities of both liquid methane and ethane systems were chosen as the average values of NPT runs at atmospheric pressure. NVT equilibration and production runs of few $ns$ followed and the size of the systems were 512 $CH_{4}$ and 500 $CH_3-CH_3$ molecules. We note here that the {\it BBK} integrator used for Langevin dynamics exhibits pressure  fluctuations of the order of $\pm 40$ {\it atm} in the liquid phase, whereas temperature fluctuations have small variance and the system is driven to the target temperature a lot faster than with conventional MD.

In order to compute the effective non-bonded coarse-grained potential, different simulation runs have been used which are discussed below.
 
\subsubsection{Constrained runs}\label{constr_runs}
The first method which we use in order to estimate the effective CG potential is by constraining the intermolecular distance between two molecules, $r=r_{1,2}$, in order to compute the constrained partition function \eqref{cg}. 
We call it ``constrained run'' of two methane, or ethane, molecules and special care had to be taken in order to avoid long sampling of the low probability short distances. 
This method is very similar to the typical conditional reversible work methods in which CG degrees of freedom are constrained at a fixed values for deriving CG potentials, as well as in free energy calculations.
Technically, we pin the centres of mass (COM) of each CG particle in space and, on every step throughout the Langevin dynamics trajectory, we subtract the total force acting on each COM.
Hence, we allow the atoms to move, resulting in rotations but not translations 
of the CG degrees of freedom ($\textrm{CH}_4$, COM).
During these runs the constraint forces are recorded. 
The mean value $\langle f \rangle_{r_{12}=r}$ is calculated in the same manner and we get $W^{(2), \text{full, f}}(r)$, from $f =-\nabla W$ . Both $W^{(2), \text{full, f}}(r)$ and $W^{(2), \text{full, u}}(r)$ are based on the same trajectory.
Then, the effective potential is calculated by numerical integration of the constraint force $\langle f \rangle_{r_{12}=r}$ from $r_{min}$ up to $r_{max}$. 

The constrained run technique described above, accelerates the sampling for short distances but there is a caveat; the ensemble average at very short distances (left part of the potential well) is strongly affected by the non-bonded forces on specific atoms between the two molecules. For example, the two $CH_{4}$ molecules are oriented according to the highly repulsive forces and rotate around the axis connecting the two COM's. Due to this specific reason, we utilized stochastic (Langevin) dynamics in order to better explore the subspace of the phase space, as a random kick breaks this alignment. We determine the minimum amount of steps needed for the ensemble average to converge, in a semi-empirical manner upon inspection of the error-bars. 

\subsubsection{Geometric direct computation of PMF}\label{geom_ave}
In order to further accelerate the sampling and alleviate the noise problems at high energy regions, that might become catastrophic in the case of the non-symmetric $CH_{3}-CH_{3}$ model, we have also calculated the two-body PMF (constraint partition function) directly, through ``full sampling" of all possible configurations using a geometrical method proper for rigid bodies. 
In more detail, the geometric averaged constrained two-body effective potential ${W}^{(2), geom}(r)$, is obtained by rotating the two $CH_4$ molecules around their COM's, through their Eulerian angles and taking account of all the possible (up to a degree of angle discretization) orientations.
The main idea is to cover every possible (discretized) orientation and associate it with a corresponding weight. The Euler angles proved to be the easiest way to implement this; each possible orientation is calculated via a rotation matrix using three (Euler) angles in spherical coordinates. 

The above way of sampling is more accurate (less noisy) than constrained MD and considerably faster. In addition, the nature of the computations allows massive parallelization of the procedure. We used a ZYZ rotation with $d\phi = d\psi= d\theta = \pi/20$ for $CH_{4}$ and simple spherical coordinate sampling with $d\phi = \pi/20, d\theta = \pi/45$ for $CH_{3}-CH_{3}$ (as it is diagonally symmetric in the united atom description).
Note however, that in this case the molecules are treated as rigid bodies; i.e., bond lengths and bond angles are kept fixed, essentially it is assumed that intra-molecular degrees of freedom do not affect the intermolecular (non-bonded potential) ones.
The advantage of this method is that we avoid long (and more expensive) molecular simulations of the canonical ensemble, which might also  get trapped in local minima and inadequately sample the phase space. 
We should also state that this method is very similar to the one used by McCoy and Curro in order to develop a $CH_4$ united-atom model from all-atom configurations.~\cite{McCoy1998}



All atomistic and coarse-grained simulations have been performed using a home-made simulation package, whereas all analysis has been executed through home-made codes in Matlab and Python.


\begin{figure}[h]
	\includegraphics[height=6em]{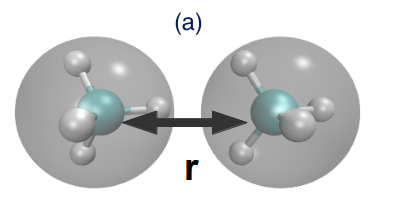}
	\includegraphics[height=6em]{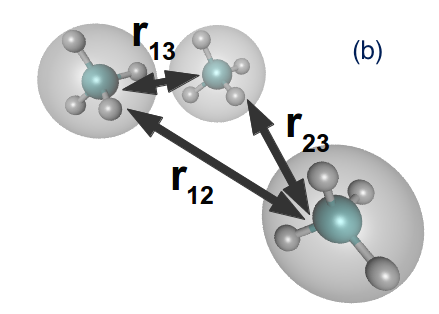}
	\includegraphics[height=20em]{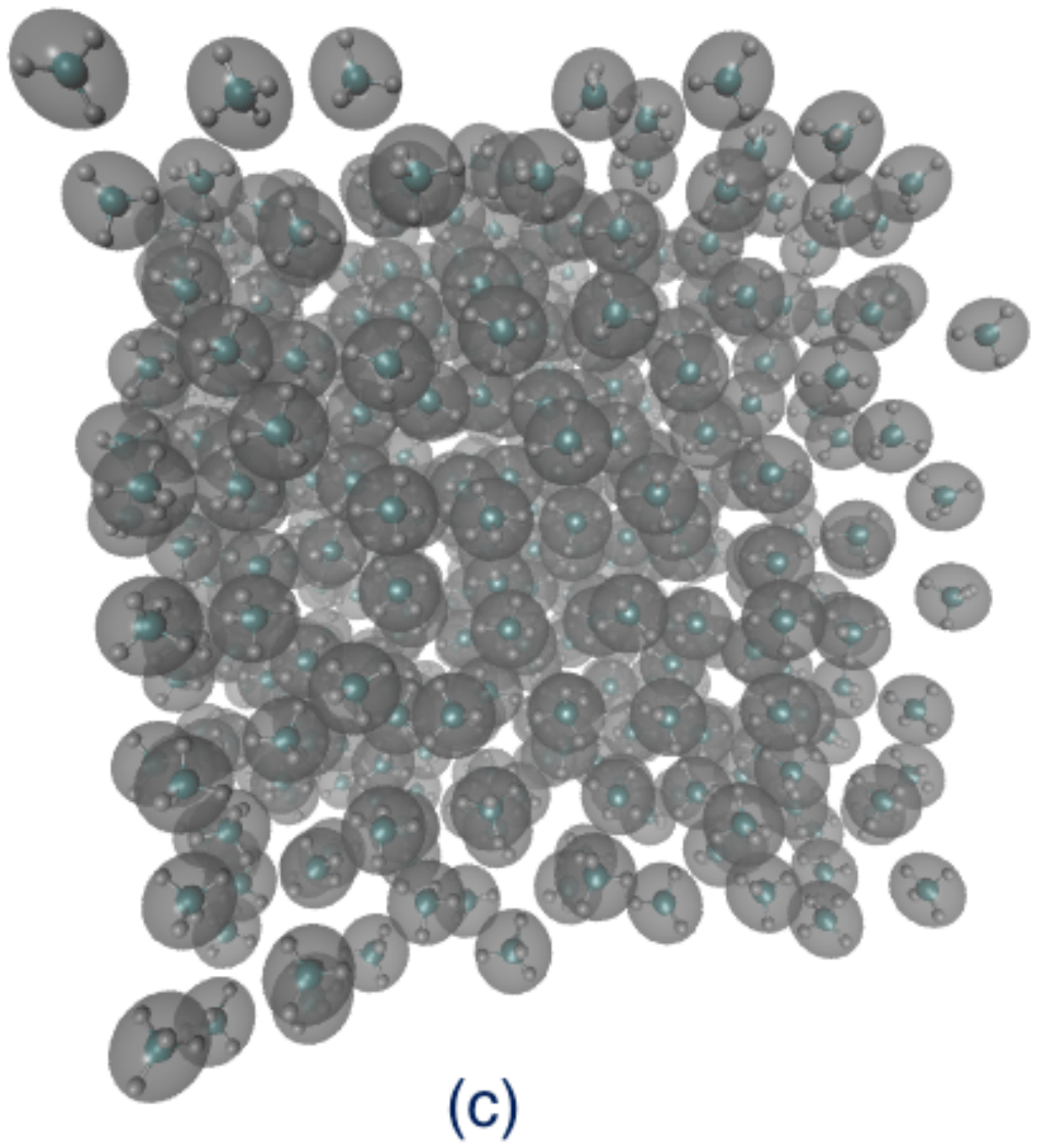}
	\caption{Snapshot of model systems in atomistic and coarse-grained description. (a-b) Two and three methanes used for the estimation of the CG effective potential from isolated molecules. (c) Bulk methane liquid.}
	\label{fig1}
\end{figure}

\section{Results}\label{sec:results}

\subsection{Calculation of the effective two-body CG potential}

\begin{figure}[ht]
	\begin{center}
	\includegraphics[height=17em]{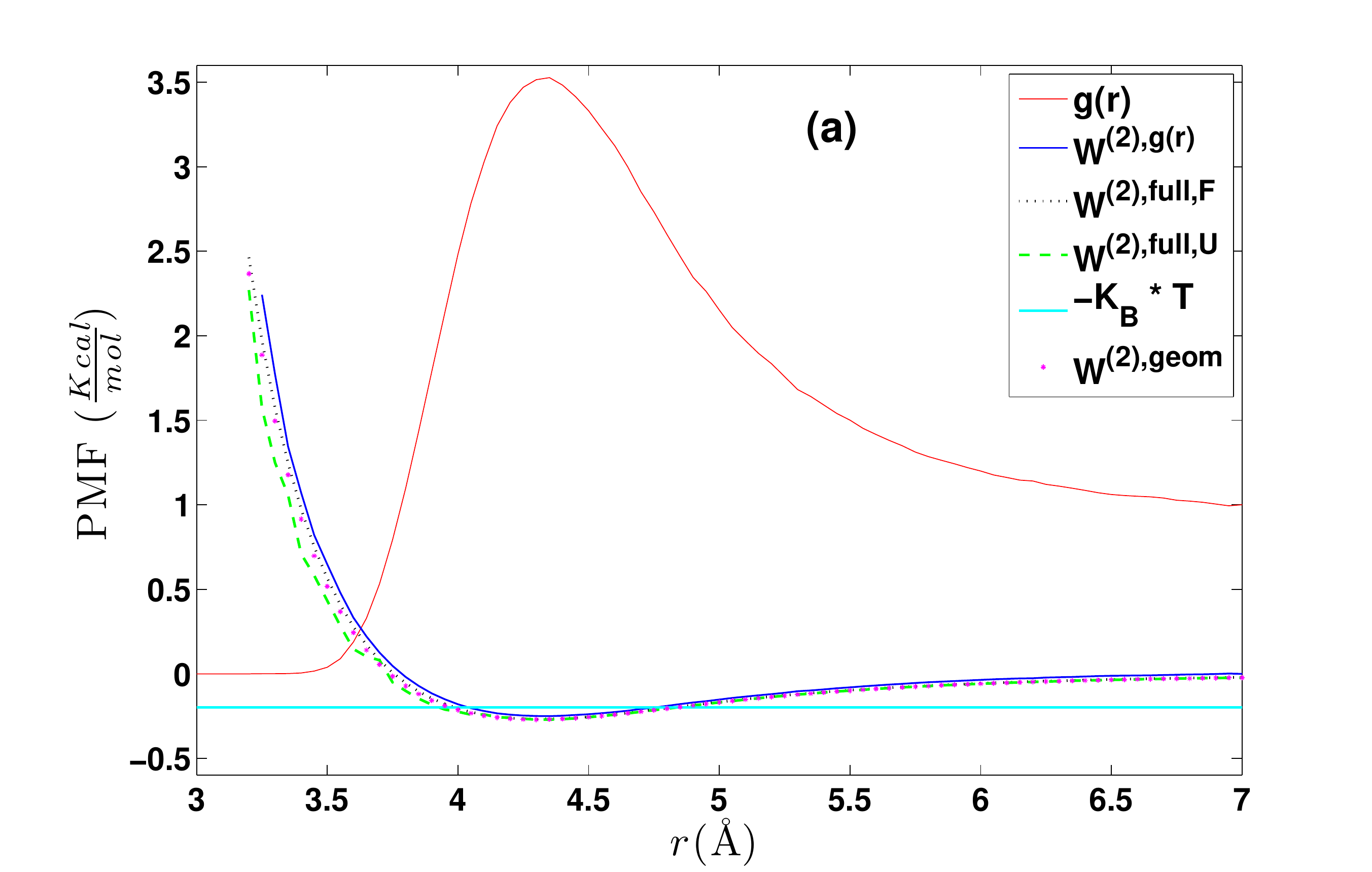}
	\includegraphics[height=17em]{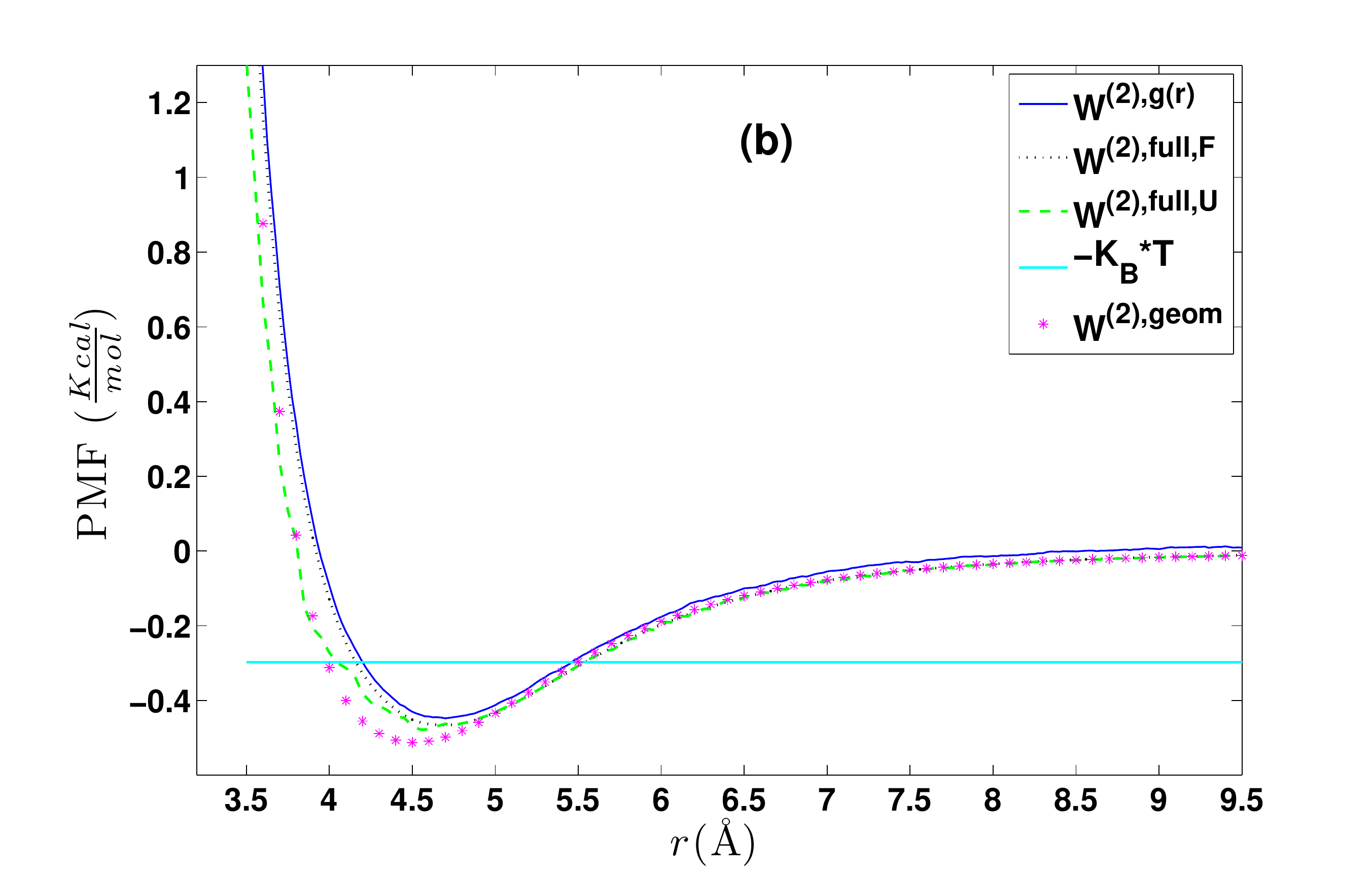}
	\end{center}
	\caption{Representation of the two-body PMF, for two isolated molecules, as a function of distance $r$, through different approximations: geometric averaging, (constrained) force matching and inversion of $g(r)$. (a) $CH_{4}$ at $T=100K$, (b) $CH_{3}-CH_{3}$ at $T=150K$. For the methane the corresponding $g(r)$ curve is also shown.}
	\label{fig:fig3}
\end{figure}

First, we present data related to the calculation of the two-body potential of mean force for the ideal system of two (isolated) molecules. For such a system the conditional M-body CG PMF is a 2-body one, i.e., the pair approximation in the effective CG interaction is exact. 
In more detail, in Figures~\ref{fig:fig3}a and \ref{fig:fig3}b we provide data for the CG effective interaction between two methane and ethane molecules, through the following methods:

(a) A direct calculation of the PMF, $W^{(2),geom}$, using a geometrical approach as described in Section \ref{geom_ave} that involves the direct calculation of the constraint partition function, treating the two molecules as rigid bodies. Note that in this case in the all-atom description bond lengths and bond angles are kept fixed. 

(b) A calculation of the PMF using the constraint force approach, $W^{(2), \text{full, f}}$, as described in section \ref{constr_runs}. In this case the constraint force required to keep  two methane molecules fixed at a specific distance is computed. Then through a numerical integration the effective potential between the two molecules (CG particles), $ U^{\text{PMF}}_{\text{CF}}$, is computed. This is a method that has been extensively used in the literature to estimate effective pair CG interaction between two molecules, as well as differences in the free energy between two states. Alternatively, through the same set of atomistic configurations the two-body PMF, $W^{(2), \text{full, u}}$, can be directly calculated through Eq~\eqref{MC2}.

(c) DBI method: The CG effective potential, $W^{(2), g(r)}$, is obtained by inverting the pair (radial) correlation function, $g(r)$, computed through a stochastic LD run with only two methane (or ethane) molecules in the simulation box. The pair correlation function, $g(r)$, of the two methane molecules is also shown in Figure~\ref{fig:fig3}a.  

The first two of the above methods refer to the direct calculation of the constrained partition function \eqref{cg} with constrained forces and canonical sampling, while the third uses the ``Direct Boltzmann Inversion" approach. All above data correspond to temperatures in which both methane and ethane are liquid at atmospheric pressure (values of $-k_BT$ are also shown in Figure~\ref{fig:fig3}). 

First, for the case of the two methane molecules (Figure~\ref{fig:fig3}a) we see very good agreement between the different methods. As expected, slightly more noisy is the $W^{(2), \text{full, u}}(r_{12})$ curve as fluctuations in the $\langle e^{-\beta u} \rangle$ term for a given $r_{12}$ distance in equation \eqref{MC2}, are difficult to cancel out. The small probability configurations in high potential energy $u$ regimes having a large impact in the average containing the exponent, hence the corresponding plot is not as smooth as the others are. In addition, as previously mentioned, $W^{(2), \text{full, f}}$ comes from the same trajectory (run) but the integration of the $\langle f \rangle_{r_{12}}$ from $r_{\text{cutoff}}$ up to $r_{12}$ washes out any non-smoothness. 
Note, that for the same system recently CG effective potentials based on IBI, force matching and relative entropy methods have been derived and compared against each other.~\cite{KCTKPH2016}

Second, for the case of the two ethane molecules (Figure~\ref{fig:fig3}b) we see a good, but not perfect, agreement between the different sets of data, especially in the regions of high potential energy (short distances).  
This is not surprising if we consider that high energy data from any simulation technique that samples the canonical ensemble, exhibit large error bars, due to difficulties in sampling. 
The latter is more important for ethane compared to methane case due to its molecular structure; indeed the atomistic structure of methane approximates much better the spherical structure of CG particles than ethane.
The only method that provides a ``full", within the numerical discretization, sampling at any distance is the geometric one; however as discussed before (see Section \ref{sec:simul}) such a method neglects the bond lengths and bond angle fluctuations. 

\begin{figure}
	\begin{center}
	\includegraphics[height=16em]{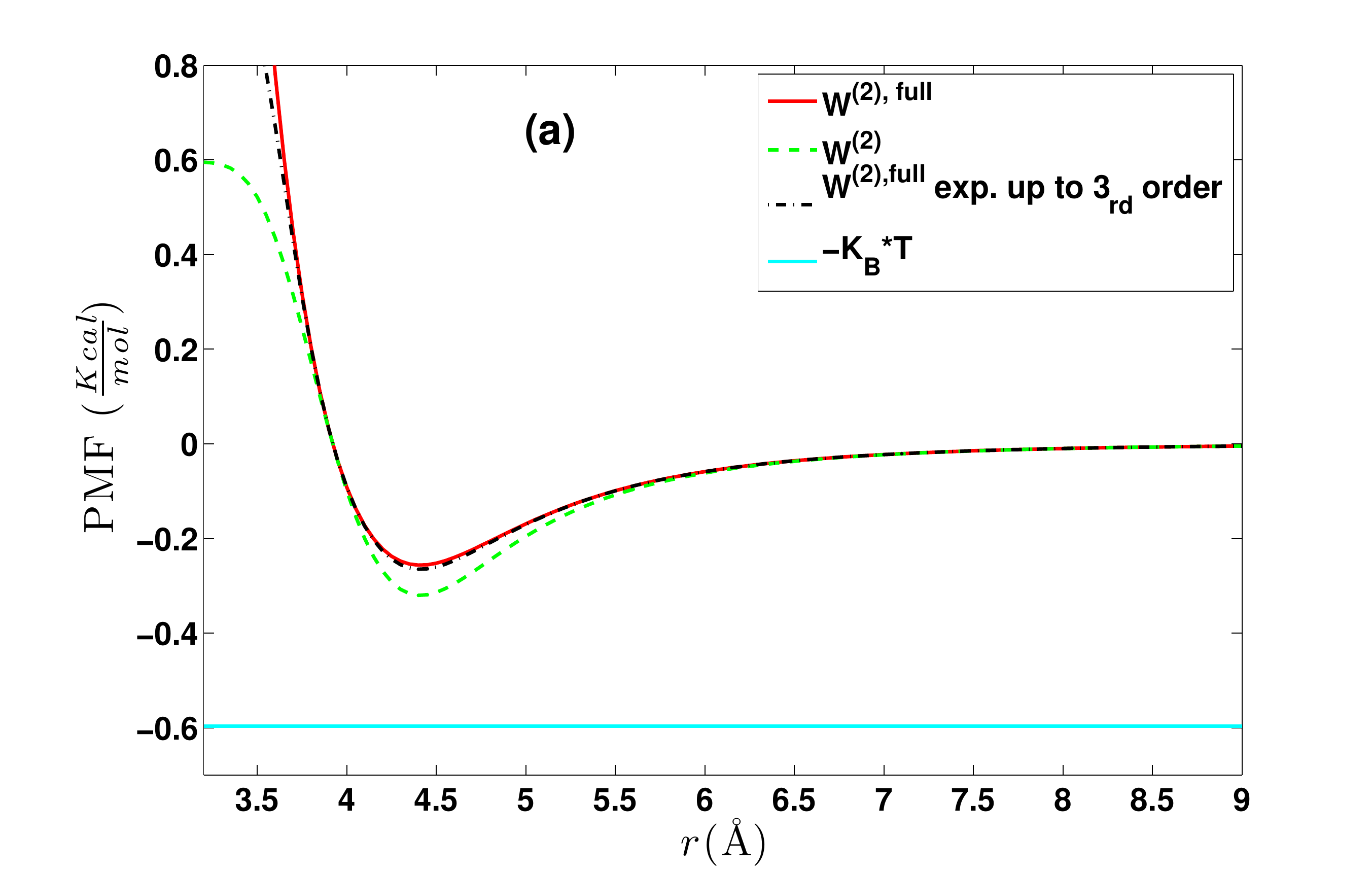}
	\includegraphics[height=16em]{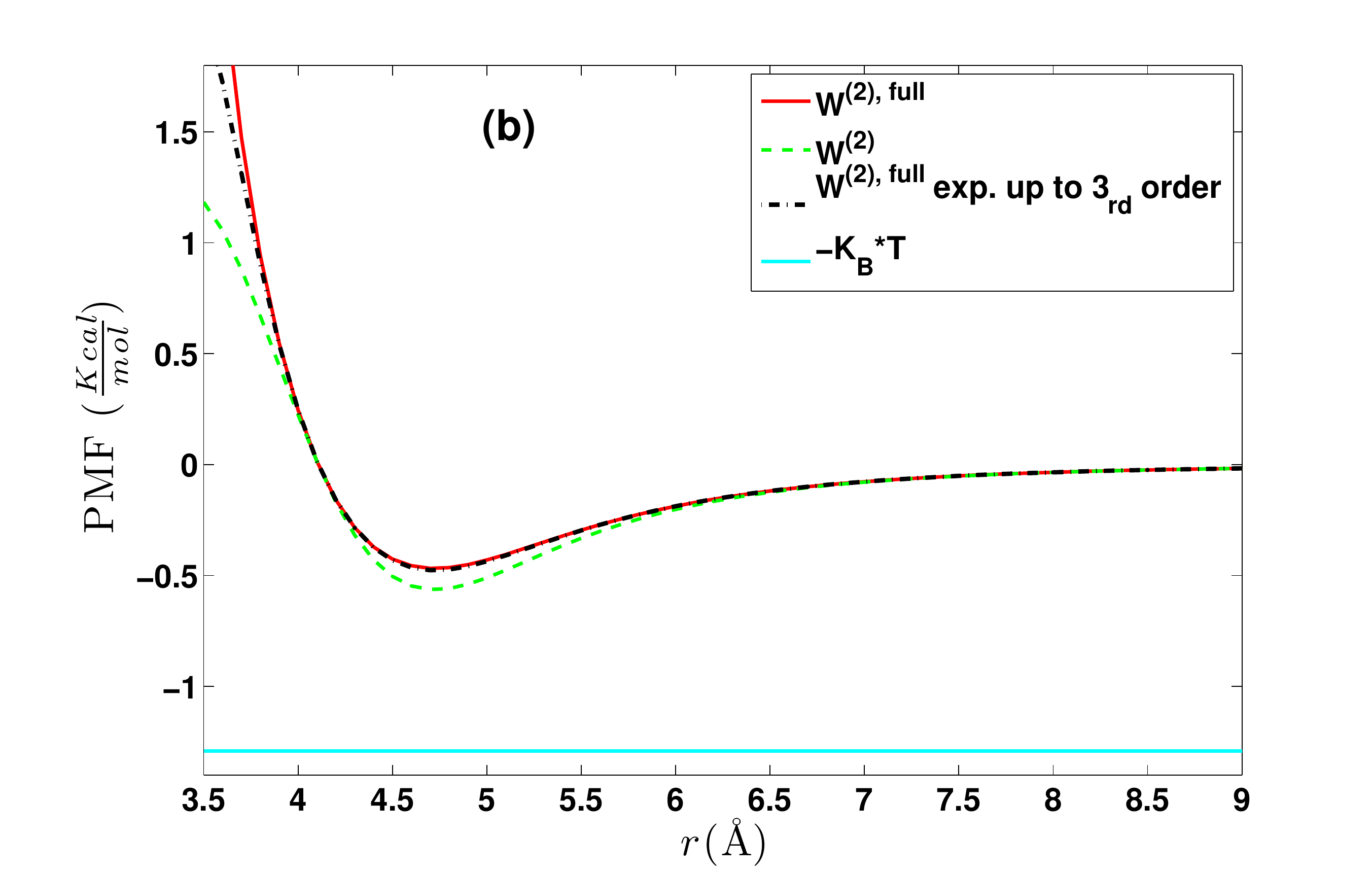}
	\end{center}
	\caption{Relation of the PMF through cluster expansions and energy averaging at high temperatures, i.e., $W^{(2)}(r_{1}, r_{2})$ and $W^{(2),\text{full}}(r_{1}, r_{2})$ through expansion over $\beta$ for $CH_{4}$ at $T=300K$ (left panel) and $CH_{3}-CH_{3}$ at $T=650K$ (right panel). As expected from the analytic form and the relation between the two formulas, $W^{(2)}$ and $W^{(2),\text{full}}$ tend to converge to the same effective potential.}
	\label{fig:fig4}
\end{figure} 

Next, we also examine an alternative method for the computation of the effective CG potential, by calculating the approximate terms from the cluster expansion approach.
For the latter we use the data from the constraint runs of two methane molecules integrated over all atomistic degrees of freedom, as given in formula \eqref{two-body}.
In Figures~\ref{fig:fig4}a and b we demonstrate the Potential of Mean Force through cluster expansions and the effect of higher order terms as shown in equation \eqref{alt}, of the two isolated molecules, for $CH_{4}$ and $CH_{3}-CH_{3}$ respectively. 
As discussed in the Section~\ref{sec:cluster-exp} cluster expansion is expected to be more accurate at high temperatures and/or lower densities. For this, we examine both systems at higher temperatures, than of the data shown in Figure~\ref{fig:fig3}; Values of $-k_BT$ are shown with full lines.
Both systems show the same behavior. First, it is clear that the agreement between $W^{(2)}$ and the (more accurate) $W^{(2),\text{full}}$ is very good only to long distances, whereas there are strong discrepancies in the regions where the potential is minimum as well as in the high energy regions (short distances).
Second, it is evident that adding terms up to the second order with respect to $\beta$, we obtain a better approximation of $W^{(2),\text{full}}$. 


\subsubsection{Effect of temperature-density}

\begin{figure}
	\begin{center}
	\includegraphics[height=17em]{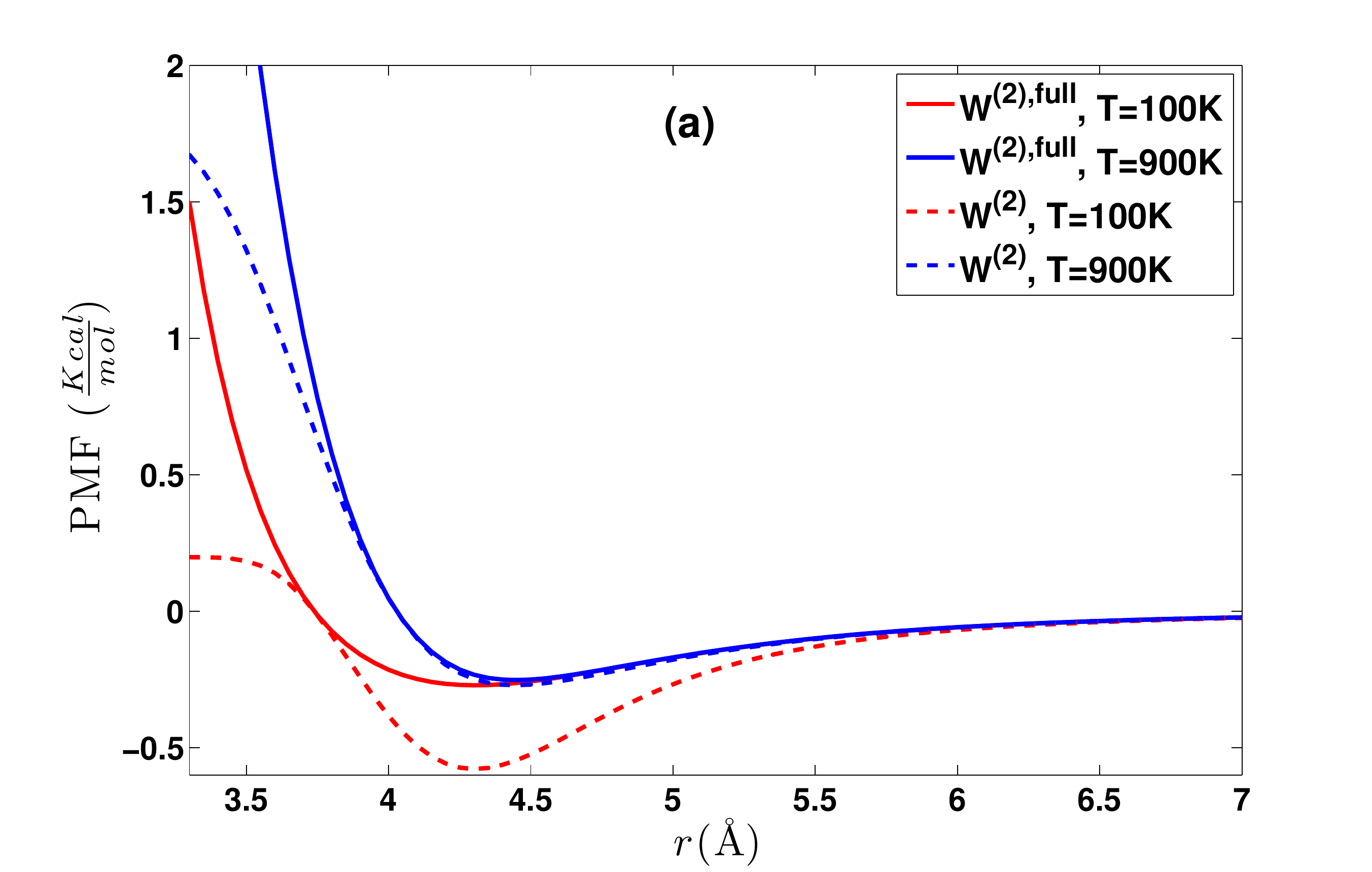}
	\includegraphics[height=17em]{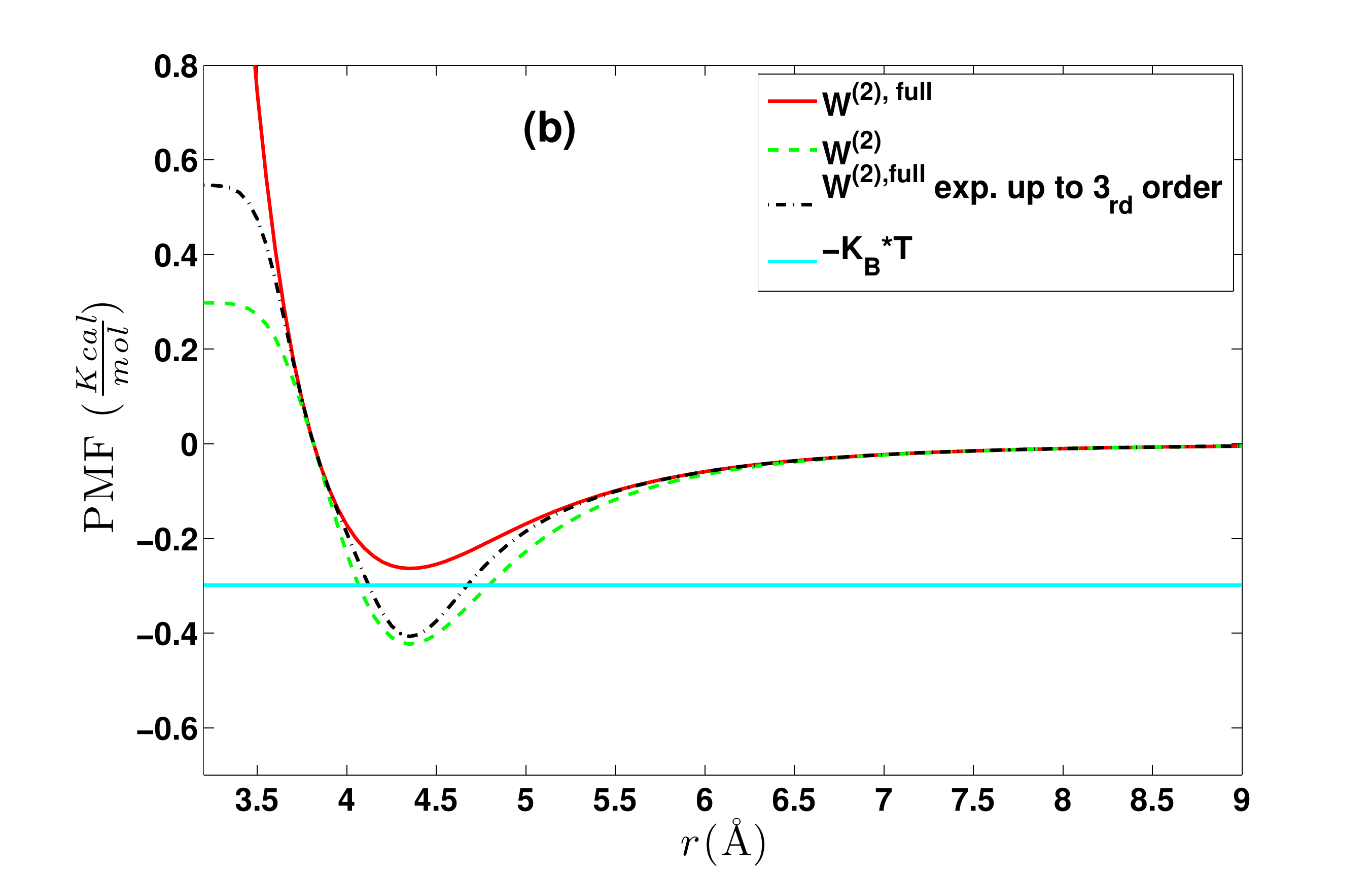}
	\end{center}
	\caption{(a) PMF through cluster expansions, using \eqref{two-body} and \eqref{alt} for different temperatures for the $CH_{4}$ model. (b) PMF through cluster expansions and energy averaging, i.e., $W^{(2)}(r_{1}, r_{2})$ and $W^{(2),\text{full}}(r_{1}, r_{2})$ through expansion over $\beta$ for $CH_{4}$ at $T=150K$.}
	\label{fig:fig5}
\end{figure} 

Next, we further examine the dependence of the PMF, for the two isolated methanes, on the temperature, by studying the molecules at $T=80K$, $120K$, $300K$ and $900K$. 
In more detail, in Figures~\ref{fig:fig5}a and b we compare the difference between $W^{(2)}$ and $W^{(2),\text{full}}$ at different temperatures.
As discussed in Section~\ref{sec:cluster-exp}, the cluster expansion method is valid only in the high temperature regime. 
This is directly observed in Figure~\ref{fig:fig5}a; at high temperatures, $W^{(2)}$  is very close to $W^{(2),\text{full}}$, which is exact for the system consisting of two molecules.
Note the small differences at short distances, which, as also discussed in the previous subsection, are even smaller if higher order terms are included in the calculation of $W^{(2)}$; see also Figure~\ref{fig:fig4}.

On the contrary, at low temperatures there is a strong discrepancy around the potential well as shown in Figure~\ref{fig:fig5}b.
In fact, for values of $r$ close to the potential well
and for rather high values of $\beta$ the contribution to the integral \eqref{temper} is large and the latter can exceed one, rendering the expansion in \eqref{alt} not valid.
In Figure~\ref{fig:fig5}b we see that the term \eqref{two-body} is not small so the expansion \eqref{alt} is not valid.
The case for ethane is qualitatively similar.  

\begin{figure}
	\resizebox{0.6\columnwidth}{!}
	{\includegraphics{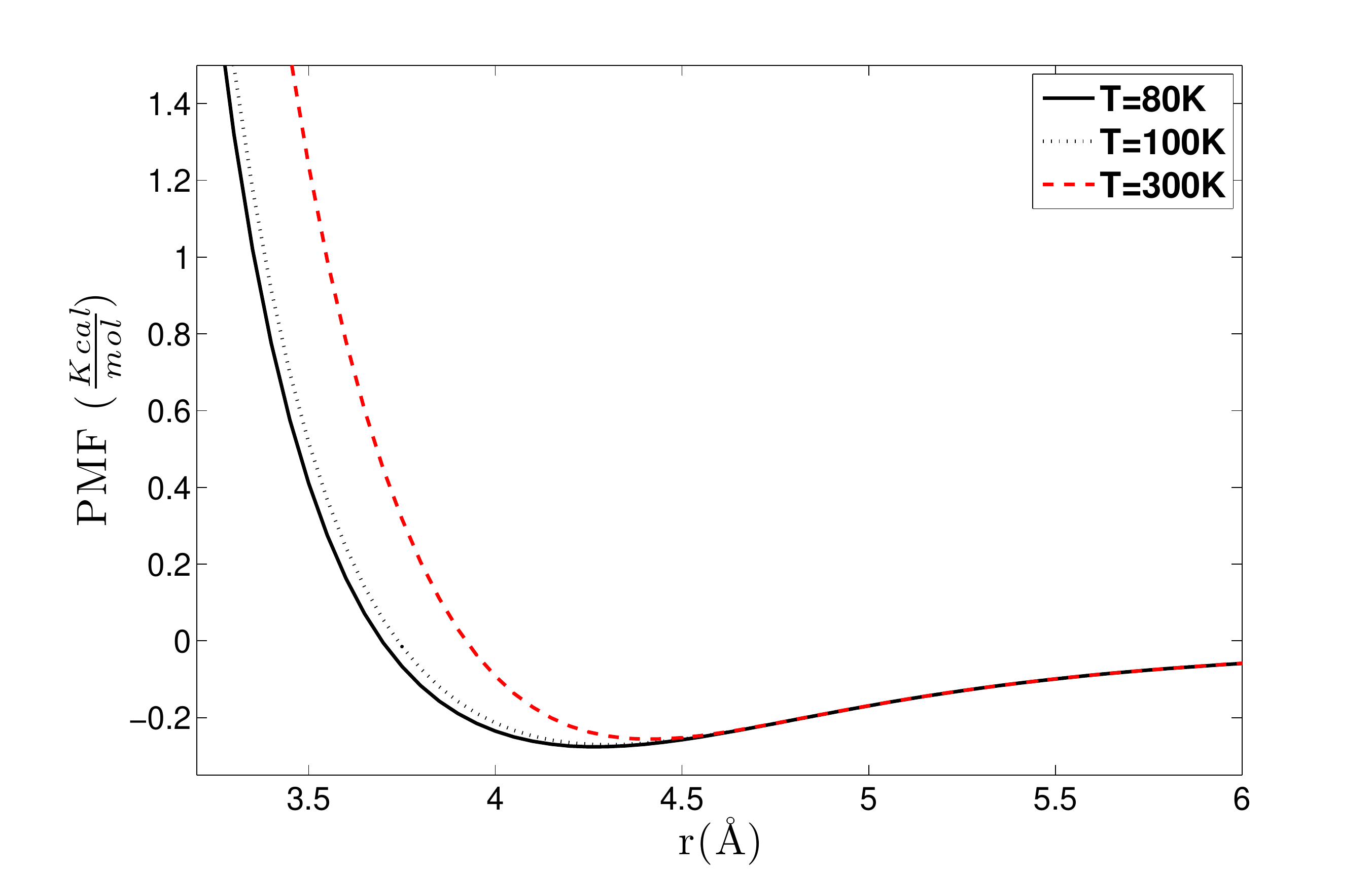}}
	\caption{Potential of mean force at different temperatures (geometric averaging). Two CH4 molecules at T=80K, 100K, 120K, 300K}
	\label{fig:fig6}
\end{figure}

For completeness, we also plot the potential of mean force at different temperatures for the system of two $CH_{4}$ molecules, see Figure~\ref{fig:fig6}. In principle, equation \eqref{two-body} is a calculation of free energy, hence it incorporates the temperature of the system and thus both approximations to the exact two-body PMF, $W^{(2)}$ and $W^{(2),\text{full}}$, are not transferable.
Indeed, we observe slight differences in the CG effective interactions (free energies) for the various temperatures, which become larger for the highest temperature. 

\subsection{Bulk CG CH4 runs using a pair potential}

In the next stage, we examine quantitatively the accuracy of the effective CG interaction potential (approximation of the two-body PMF), in the liquid state based on structural properties like $g(r)$. 
Here we use the different CG models (approximated pair CG interaction potentials) derived above, to predict the properties of the bulk CG methane and ethane liquids. 
In all cases we compare with structural data obtained from the reference all-atom bulk system, projected on the CG description.

\begin{figure}
	\includegraphics[height=17em]{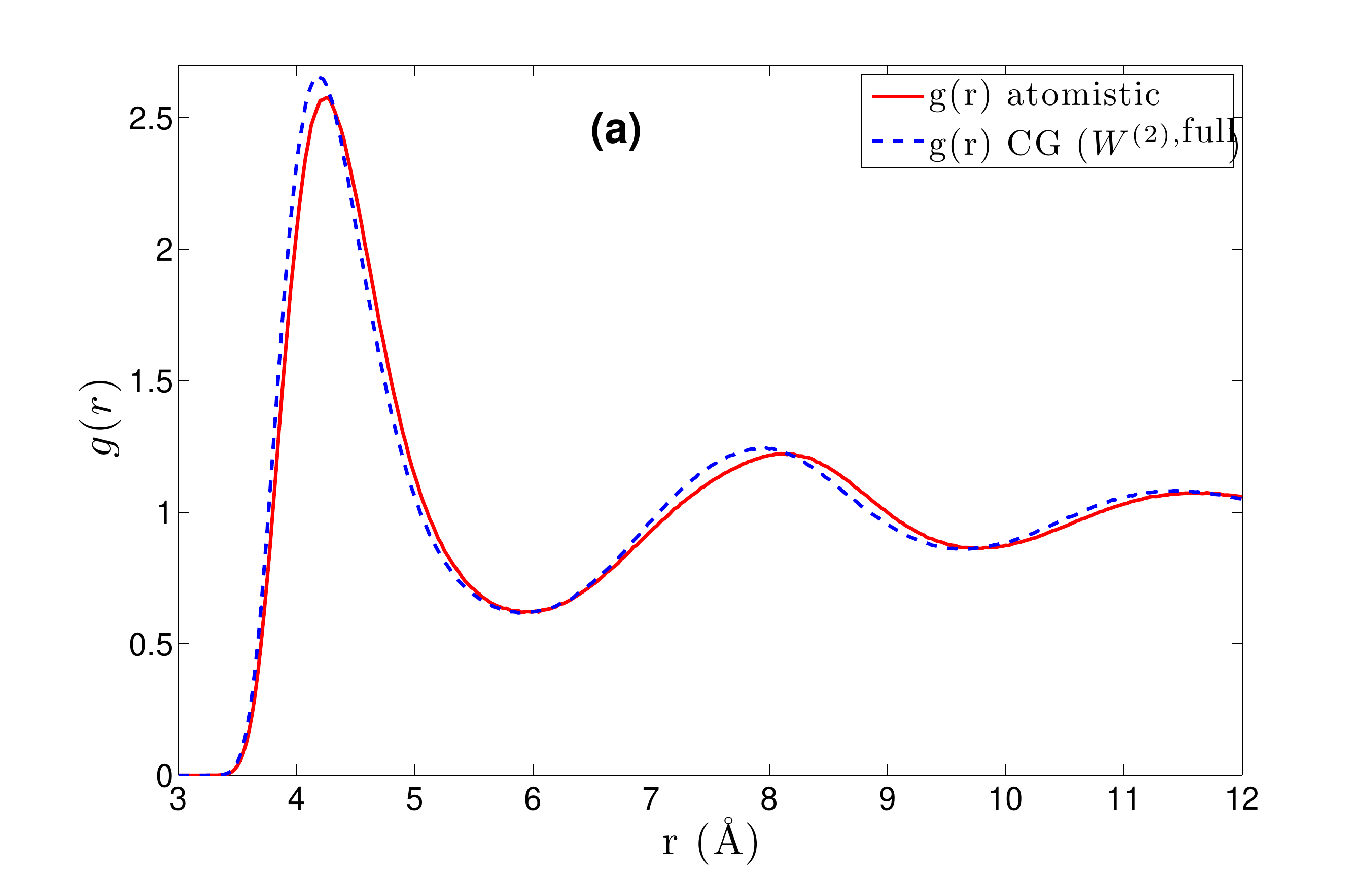}
	\includegraphics[height=17em]{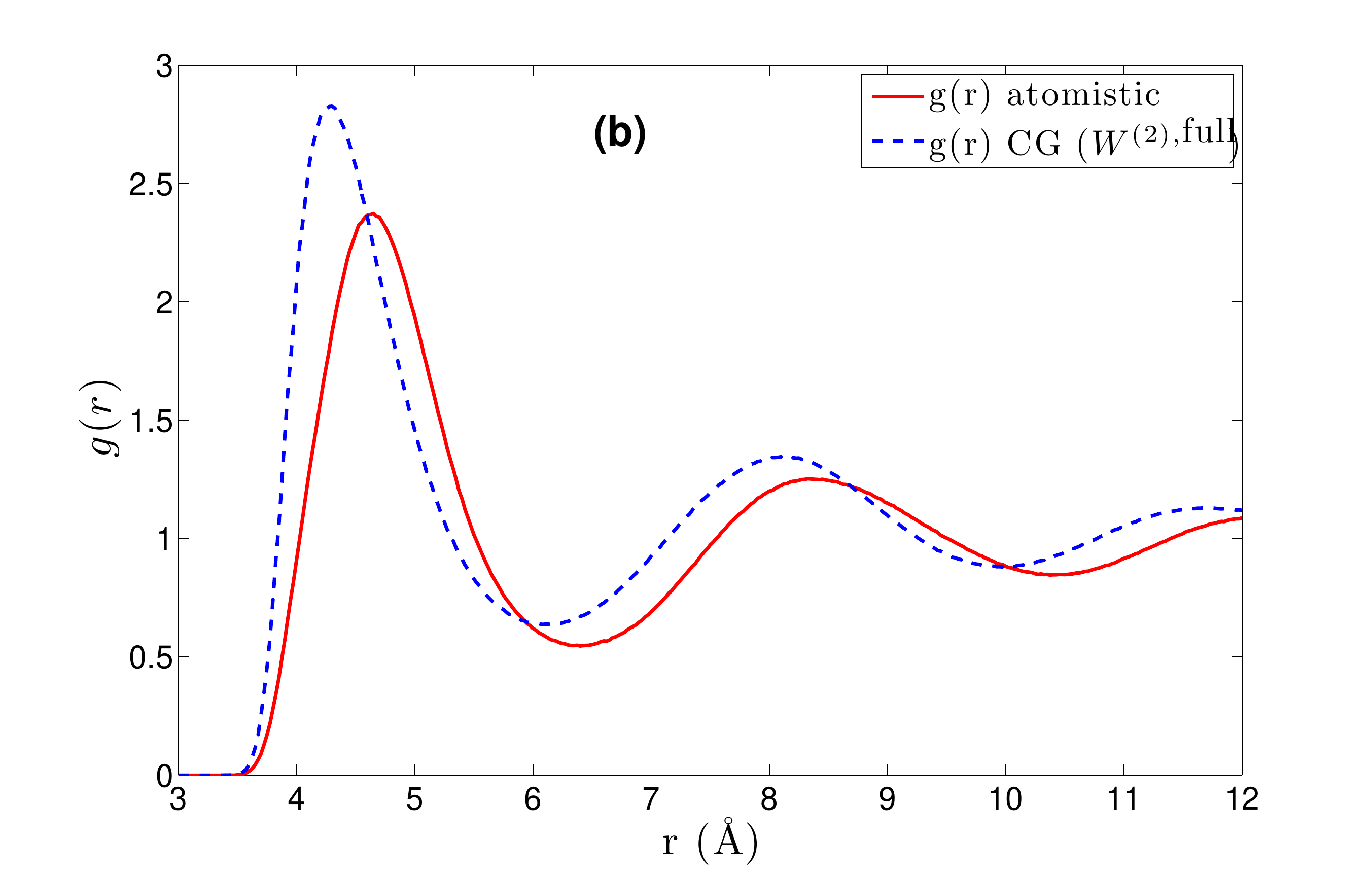}
	\caption{RDF from atomistic and CG using pair potential, $W^{(2)}$, for $CH_{4}$ system at $T=80K$ (left panel) and $CH_{3}-CH_{3}$ at $T=150$(right panel). Spherical CG approximation to the non-symmetric ethane molecule induces discrepancy implies there is more room for improvement.}
	\label{fig:fig7}
\end{figure}

In Figures~\ref{fig:fig7}a and b we assess the discrepancy between the CG (projected) pair distribution function, $g(r)$, taken from an atomistic run, and the one obtained from the corresponding CG run based on $W^{(2),\text{full}}$ as already seen in Figure~\ref{fig:fig3} of methane and ethane respectively. 
Note that $g(r)$ is directly related to the effective CG potentials  ($N=2$ in Eq~\eqref{eq:gofr_atom}). 

It is clear that for methane (Figure~\ref{fig:fig7}a) the CG model with the $W^{(2),\text{full}}$ potential gives a g(r) very close to the one derived from the analysis of the all-atom data. 
This is not surprising if we consider that for most molecular systems small differences in the interaction potential lead to even smaller differences in the obtained pair correlation function.
Interestingly the CG model with the $W^{(2)}$ is also in good agreement to the reference one, despite the small differences in the CG interaction potential discussed above (see Figures~\ref{fig:fig4} and~\ref{fig:fig8}). As expected, the difference comes from the missing higher order terms of eq~\eqref{approx}. 

The fact that the CG effective potential, which is derived from two isolated methane molecules, give a very good agreement for the methane structure in the liquid state is not surprising if we consider the geometrical structure of methane, which is rather close to the spherical one, and the typical van der Waals type of interactions between methane molecules. 
On the contrary, for the case for ethane (Figure~\ref{fig:fig7}b) predictions of $g(r)$ using pair CG potential are much different compared to the atomistic one, especially for the short distances. 
Even larger differences would be expected for more complex systems with long-range interactions, such as water.~\cite{KCTKPH2016}
Similar is the case also for the other temperatures (T = 80K) studied here (data not shown here).

\subsubsection{Effect of temperature-density}

\begin{figure}
	\begin{center}
	\includegraphics[height=17em]{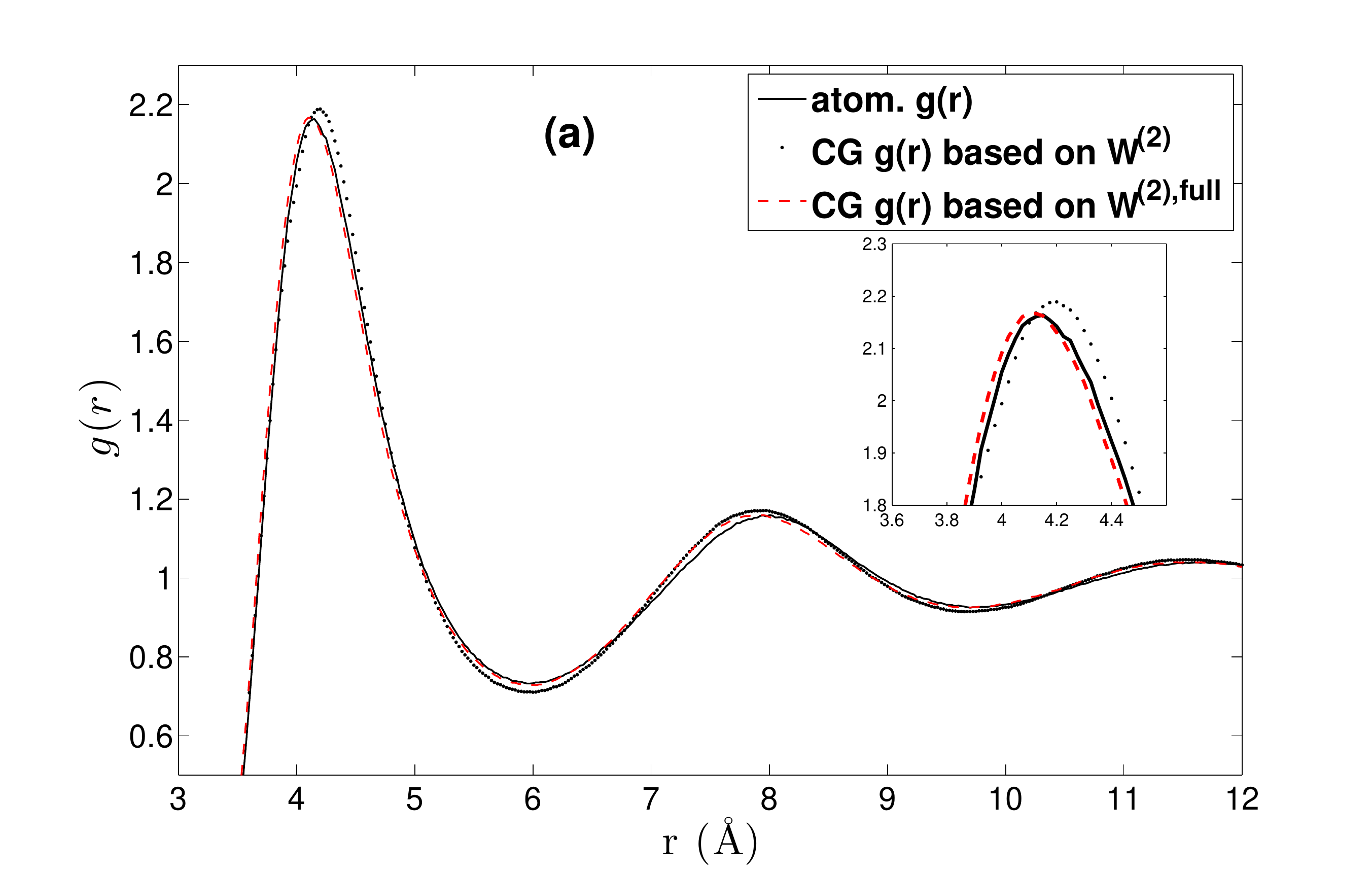}
	\includegraphics[height=17em]{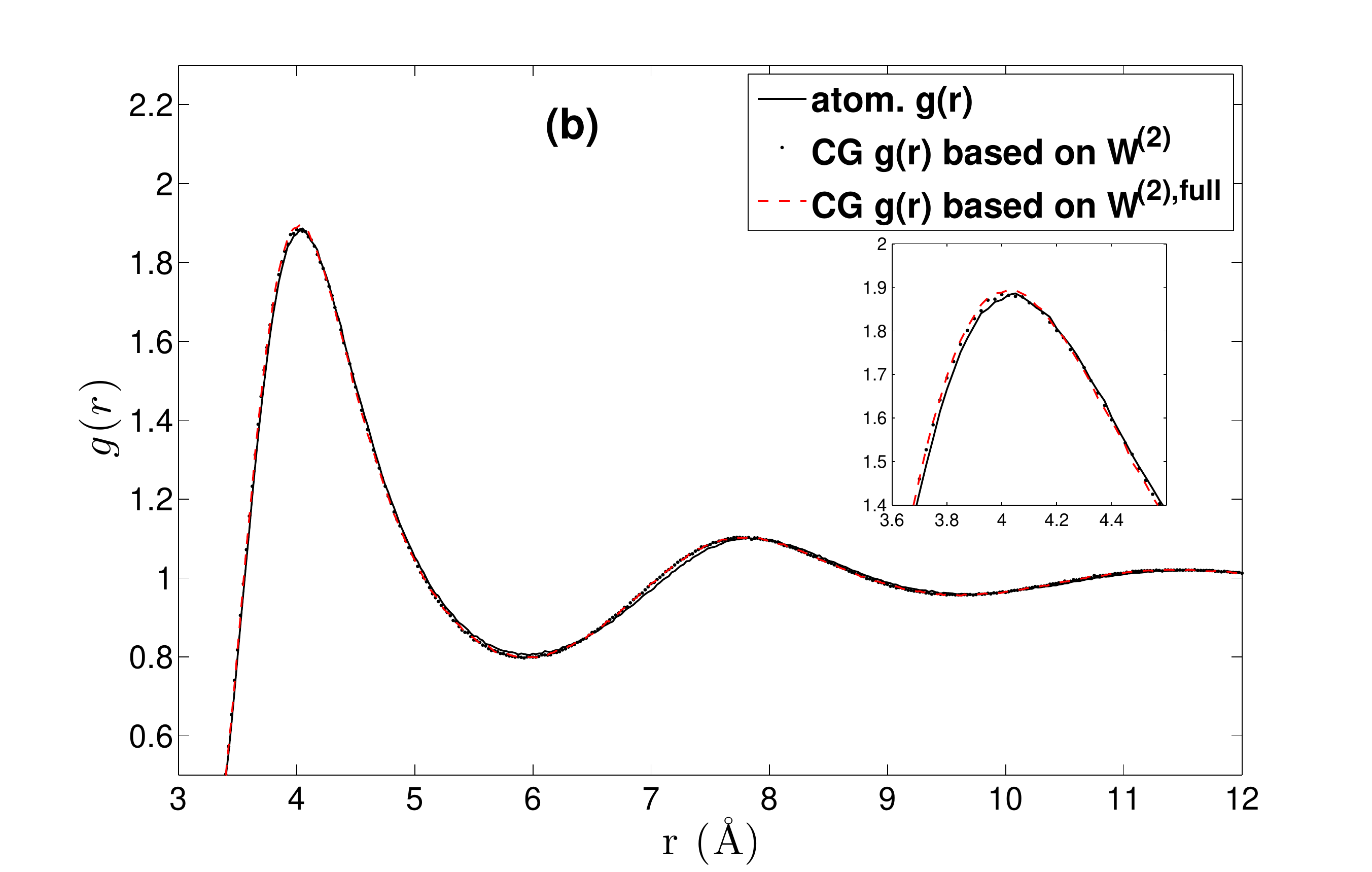}
	\end{center}
	\caption{RDF from atomistic data, and CG models using pair potential at different temperatures: (a) T=300K, (b) T=900K. In both cases the density is 0.3799 $\frac{gr}{cm^3}$.}
	\label{fig:fig8}
\end{figure}

We further study the structural behavior of the CG systems at different state points; i.e., temperature/density conditions, compared to the atomistic ones.
First, we examine the temperature effect by simulating the systems discussed above (see Figure~\ref{fig:fig7}) at higher temperatures; however keeping the same density.
In Figures \ref{fig:fig8}a,b we present the RDF of methane from atomistic and CG runs using pair potential at $T=300K$, and $T=900K$ respectively.

It is clear that the analysis of the CG runs using the $W^{(2),\text{full}}$ potential gives a pair distribution function $g(r)$ close to the atomistic one for both (high) temperatures, similar to the case of the $T=100K$ shown above. 
In addition, the CG model with the $W^{(2)}$ potential is in very good agreement with the atomistic data at high temperature (Figure \ref{fig:fig8}b), whereas there are small discrepancies at lower temperatures (Figure \ref{fig:fig8}a), in particular at the maximum of $g(r)$. This is shown in the inset of Figures~\ref{fig:fig8}a,b.
Note also that in this high temperature the incorporation of the higher order terms in $W^{(2)}$ leads to very similar potential as the $W^{(2),\text{full}}$ (see also Fig.~\ref{fig:fig4}), and consequently to very accurate structural $g(r)$ data as well.

\begin{figure}
	\begin{center}
	\includegraphics[height=17em]{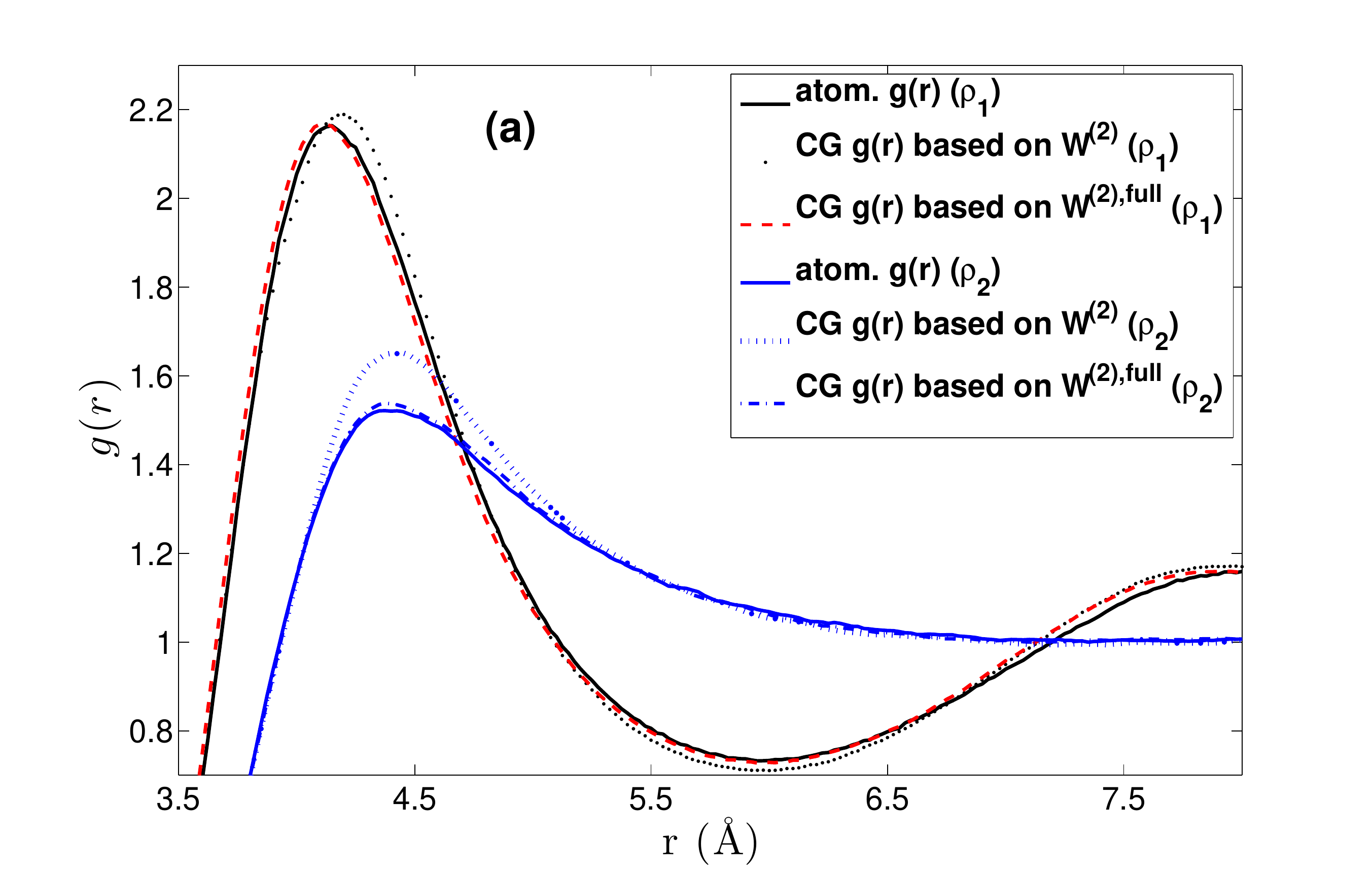}
	\includegraphics[height=17em]{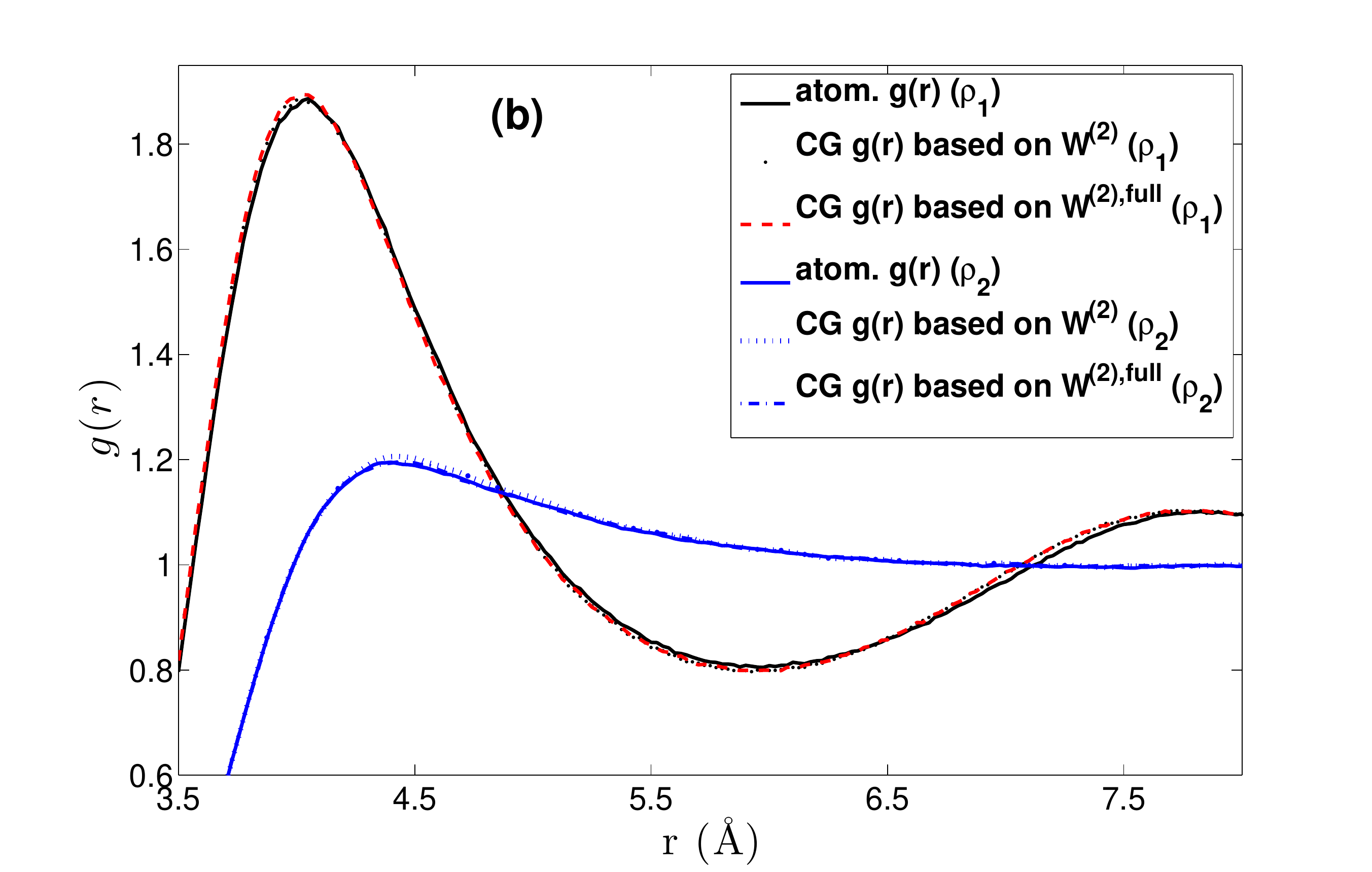}
	\end{center}
	\caption{RDF from atomistic and CG using pair potential at different densities $\rho_1<\rho_2$, (a) T=300K, (b) T=900K.}
	\label{fig:fig9}
\end{figure}

Next, we examine the structural behavior of the CG systems at different densities.
In Figure \ref{fig:fig9}a we present the $g(r)$ from atomistic and CG runs using pair potential at different densities ($\rho_1= 0.3799 \frac{gr}{cm^3}$ and $\rho_2= 0.0395 \frac{gr}{cm^3}$ and  $T=300K$, and  $T=900K$). 
There is apparent discrepancy from the reference (atomistic) system in both densities in agreement to the data discussed above in Figure \ref{fig:fig8}a.

For the case of higher temperature data ($T=900K$)
and the same densities, as shown in Figure \ref{fig:fig9}b, the pair distribution function, $g(r)$, obtained from the CG model with the $W^{(2)}$ effective interaction is very close to the data derived from the $W^{(2),\text{full}}$ one, and in very good agreement to the reference, all-atom, data.
This is not surprising since, as discussed before, at high temperatures the cluster expansion is expected to be more accurate, since cluster expansions hold for high $T$ and low $\rho$. 
From Figure~\ref{fig:fig9}a we deduce that despite the different potentials $W^{(2)}, W^{(2),\text{full}}$ (Figure~\ref{fig:fig4}), we obtain the same $g(r)$ for the liquid case, as a result of the close packing and frequent collisions.

Overall, the higher the temperature the better the agreement in the $g(r)$ derived from the CG models using any of $W^{(2)}$ and $W^{(2),\text{full}}$. These data are in better agreement with the atomistic data as well. 


\section{Effective three-body potential}

In the last part of this work we briefly discuss the direct computation of the three-body effective CG potential and its implementation in a (stochastic) dynamic simulation. More results about the three-body terms will be presented in a future work.~\cite{THT2017}

\subsection{Calculation of the effective three-body potential}

In the following we present data for the 3-body potential of mean force estimated from simulation runs and geometric computations involving three isolated molecules. We have two suggestions for the 3-body PMF: 
(a) Formula \eqref{three-body} derived from cluster expansion formalism,  which is valid for rather high temperatures and 
(b) another one based on the McCoy-Curro scheme given in formula~\eqref{MC3}.

Similarly to the two-body potential, 
the corresponding calculations can be performed by running constrained molecular dynamics (or any other method that performs canonical sampling).
For this one needs to calculate the derivative of the three-body potential with respect to some distance. However, as previously stated, deterministic MD simulations of a constrained system might easily get trapped in local energy minima, so we utilized stochastic dynamics for the three-body case. In addition, rare events (high energy, low probability configurations) induce noise to the data, despite long equilibration (burn-in) periods or stronger heat-bath coupling in the simulations. Although smoothing could in principle have been applied, it would wash-out important information needed upon derivation with respect to positions ($f = - \nabla_{\bf{q}} W^{(3)} $). Therefore, we choose here to present results from the ``direct" geometric averaging approach. The total calculations are one order of magnitude more than the two-body ones (all possible orientations of the two molecules for one of the third one), so special care was given to spatial symmetries. 

\begin{figure}
	\begin{center}
	\includegraphics[height=17em]{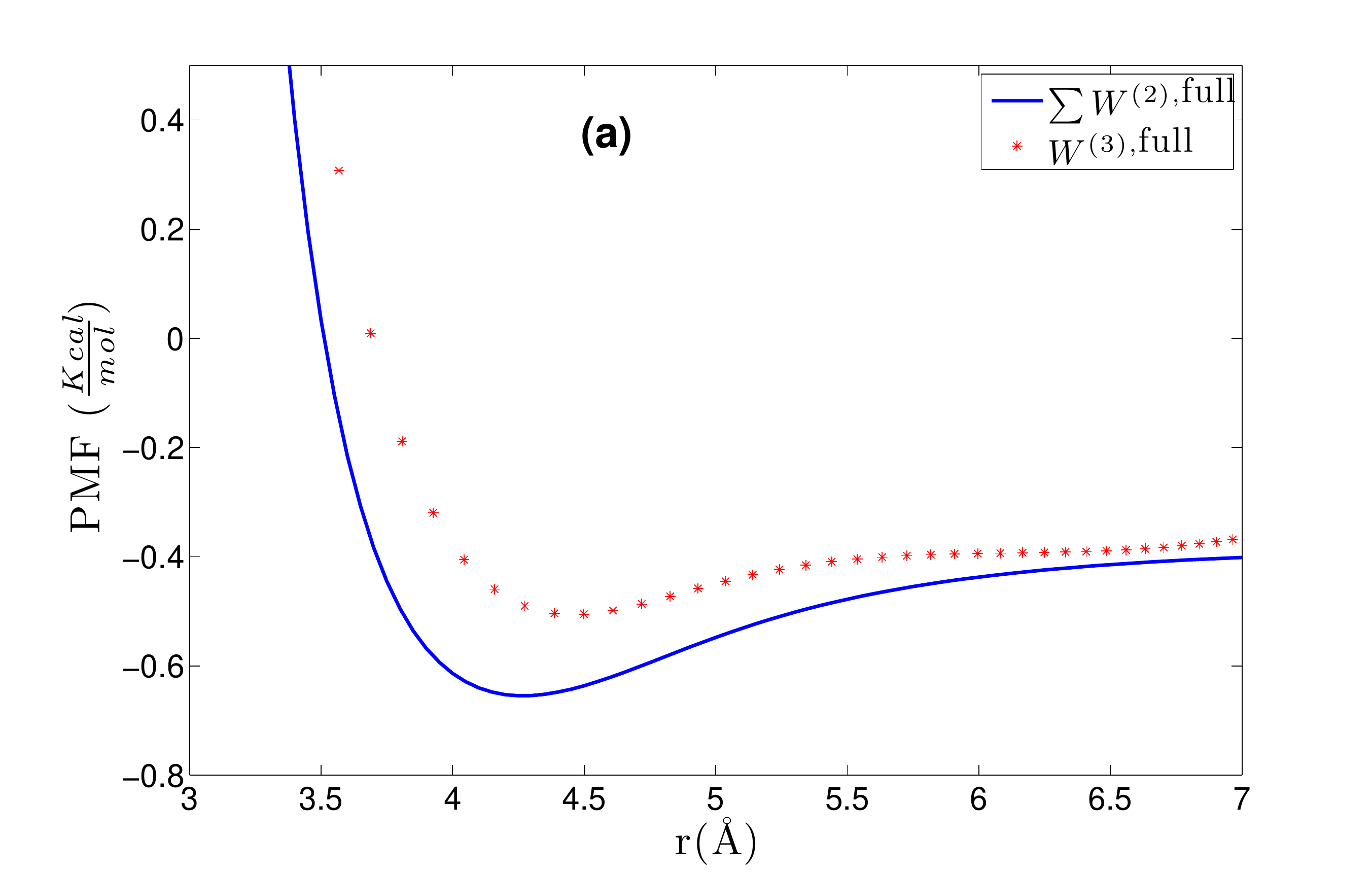}
	\includegraphics[height=17em]{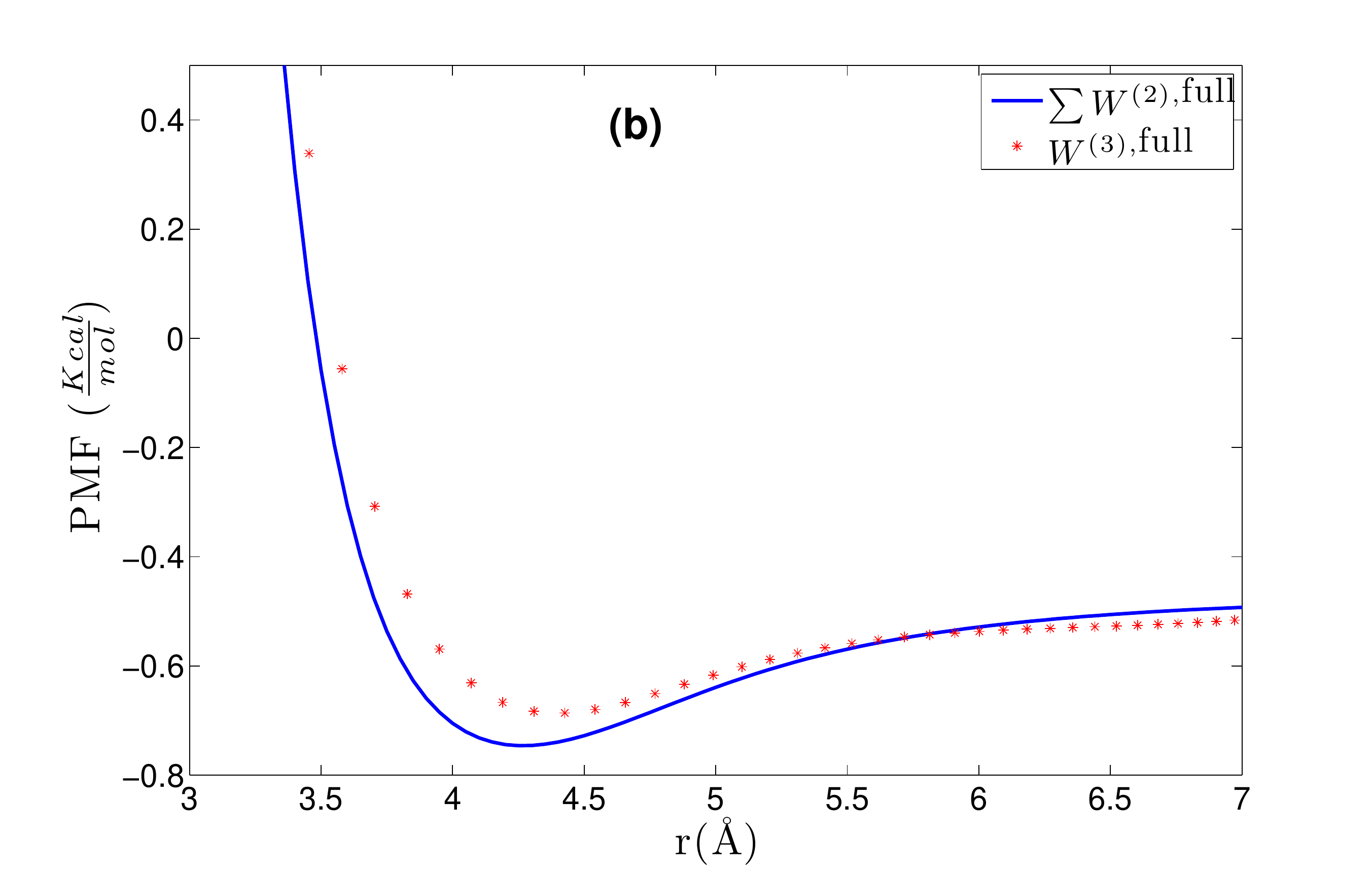}
	\includegraphics[height=17em]{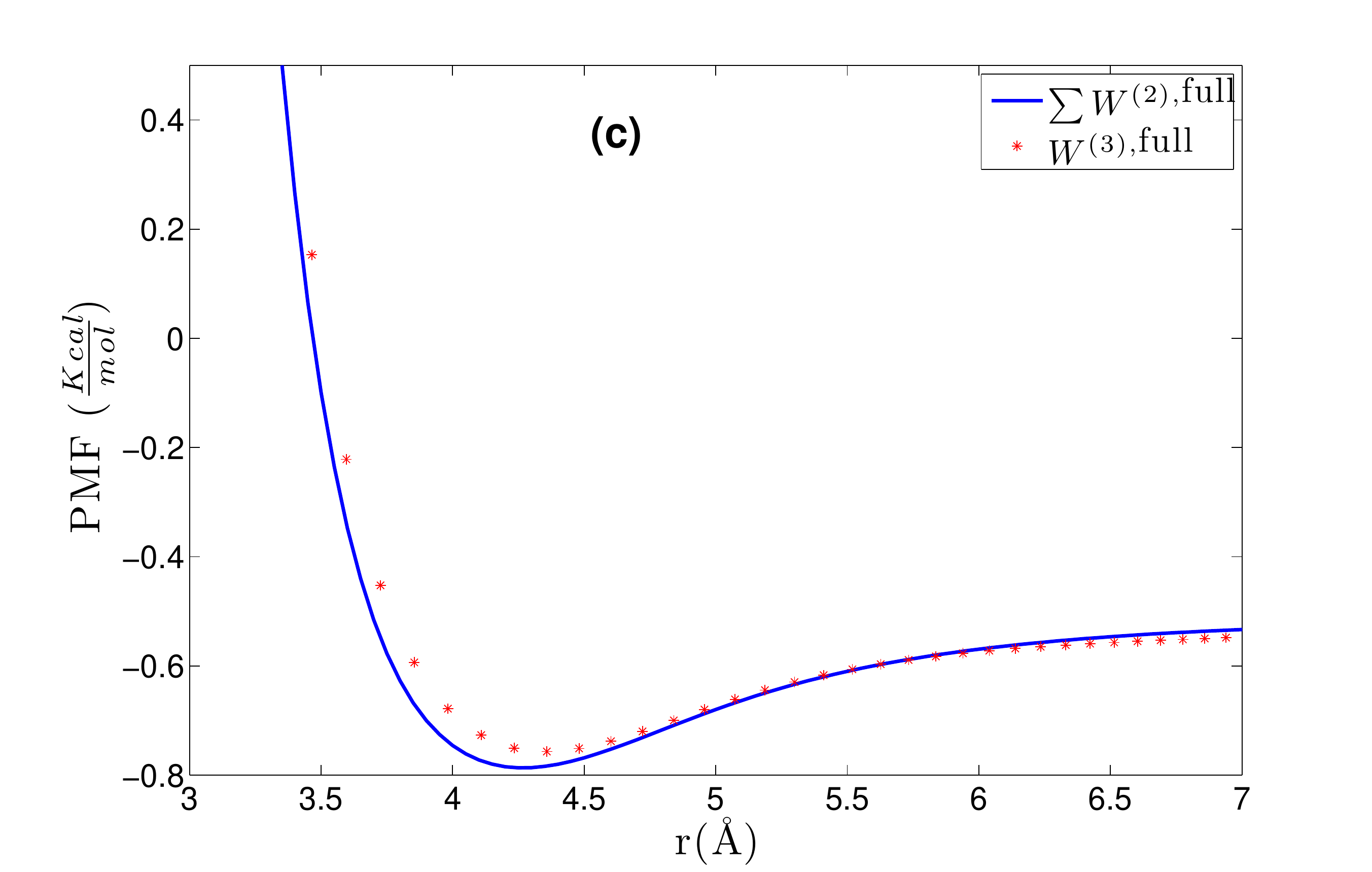}
	\includegraphics[height=17em]{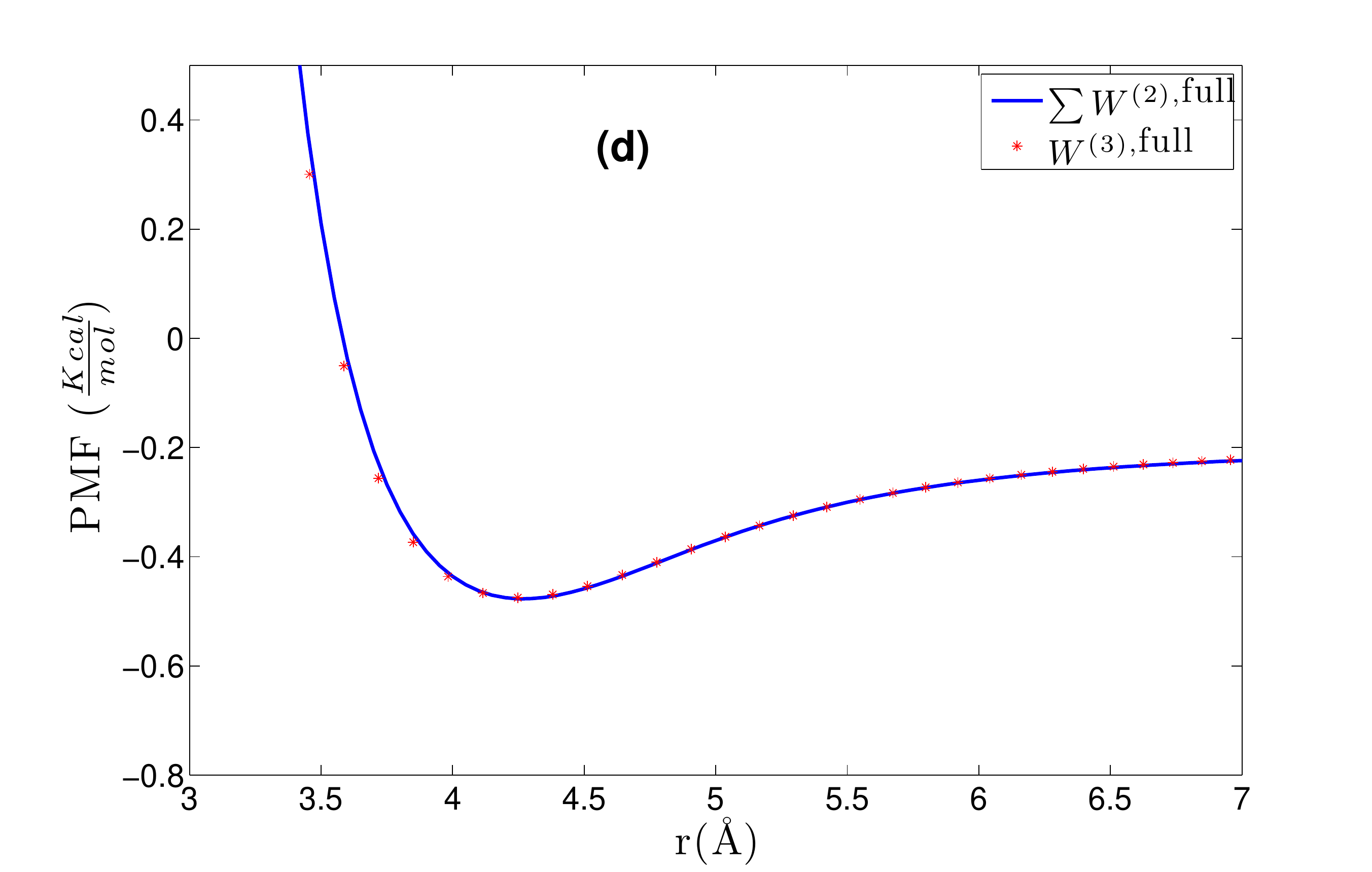}
	\end{center}
	\caption{Effective potential comparison between the $W^{(3),\text{full}}$ 3-body and $\sum W^{(2),\text{full}}$ simulations (geometric averaging) for $CH_{4}$ at $T=80K$ for different COM distances [$\AA$] . (a) $r_{12}=3.9, r_{13}=3.9$ (b)$r_{12}=4.0, r_{13}=4.0$ (c) $r_{12}=4.3, r_{13}=4.0$, (d)$r_{12}=3.8, r_{13}=5.64$ .}
	\label{fig:fig10}
\end{figure}

The new effective three-body potential, $W^{(3),\text{full}}(r_{12}, r_{13}, r_{23})$, incorporates three intermolecular distances: $r_{12}, r_{13}, r_{23}$. 
The discretization of the COM's in space is on top of the angular discretization mentioned in Section~\ref{geom_ave} and relates to the above three distances.
In more detail, in Figures \ref{fig:fig10}a-d we present simulations based on the effective three body potential $W^{(3), \text{full}}$ and the sum $\sum W^{(2),\text{full}}$ (geometric 
averaging) for $CH_{4}$ at $T=80K$ for different COM distances [$\AA$]: (a) $r_{12}=3.9$, 
$r_{13}=3.9$, (b)$r_{12}=4.0$, $r_{13}=4.0$, (c) $r_{12}=4.3$, $r_{13}=4.0$, 
(d) $r_{12}=3.8$, $r_{13}=5.64$. 
In each case the sum of the corresponding two-body terms is also shown.
At smaller distances, the potential of the triplet deviates from the sum of the three pairwise potentials and this is where improvement in accuracy can be obtained. 
As shown in Figure~\ref{fig:fig10} improvement is needed for close distances around the (3 dimensional) well. We used a 3-dimensional cubic polynomial to fit the potential data (conjugate gradient method) which means that 20 constants should be determined. A lower order polynomial cannot capture the curvature of the forces upon differentiation. The benefit of this fitting methodology (over partial derivatives for instance) is the analytical solution of the forces with respect to any of $r_{12}, r_{13}, r_{23}$ in contrast to tabulated data that induce some small error.

Overall, there are clear differences between the 3-body PMF, $W^{(3),\text{full}}$, and the sum of three two-body interactions, $\sum W^{(2),\text{full}}$, at short $r_{12}$, $r_{13}$ and $r_{23}$ distances. On the contrary, for larger distances the sum of two-body interactions seems to represent the full three-body PMF very accurately. This is a clear indication of the rather short range of the three-body terms.   
Based on the above data, the range of the 3-body terms for this system (methane at $T=80K$) is: $r_{12}\in[3.8:4.1]\AA, r_{13}\in[3.8:4.1]\AA$ and $ r_{23}\in[3.8:5] \AA$; hence, the maximum distance for which three-body terms were considered, is $r_{\text{cut-off},3}$=5$\AA$.
In practice we need to identify all possible triplets within $r_{\text{cut-off},3}$.
Naturally, by including higher-order terms the computational cost has increased as well.
More information about the numerical implementation of the three-body CG effective potential and its computational efficiency will be given elsewhere.~\cite{THT2017}

We should state here that in order to keep constant the temperature (in the BBK algorithm) due to the extra three-body terms in the CG force field a larger coupling constant value 
for the heat bath was required.

\subsection{CG Runs with the effective three-body potential}

\begin{figure}
	\resizebox{0.6\columnwidth}{!}
	{\includegraphics{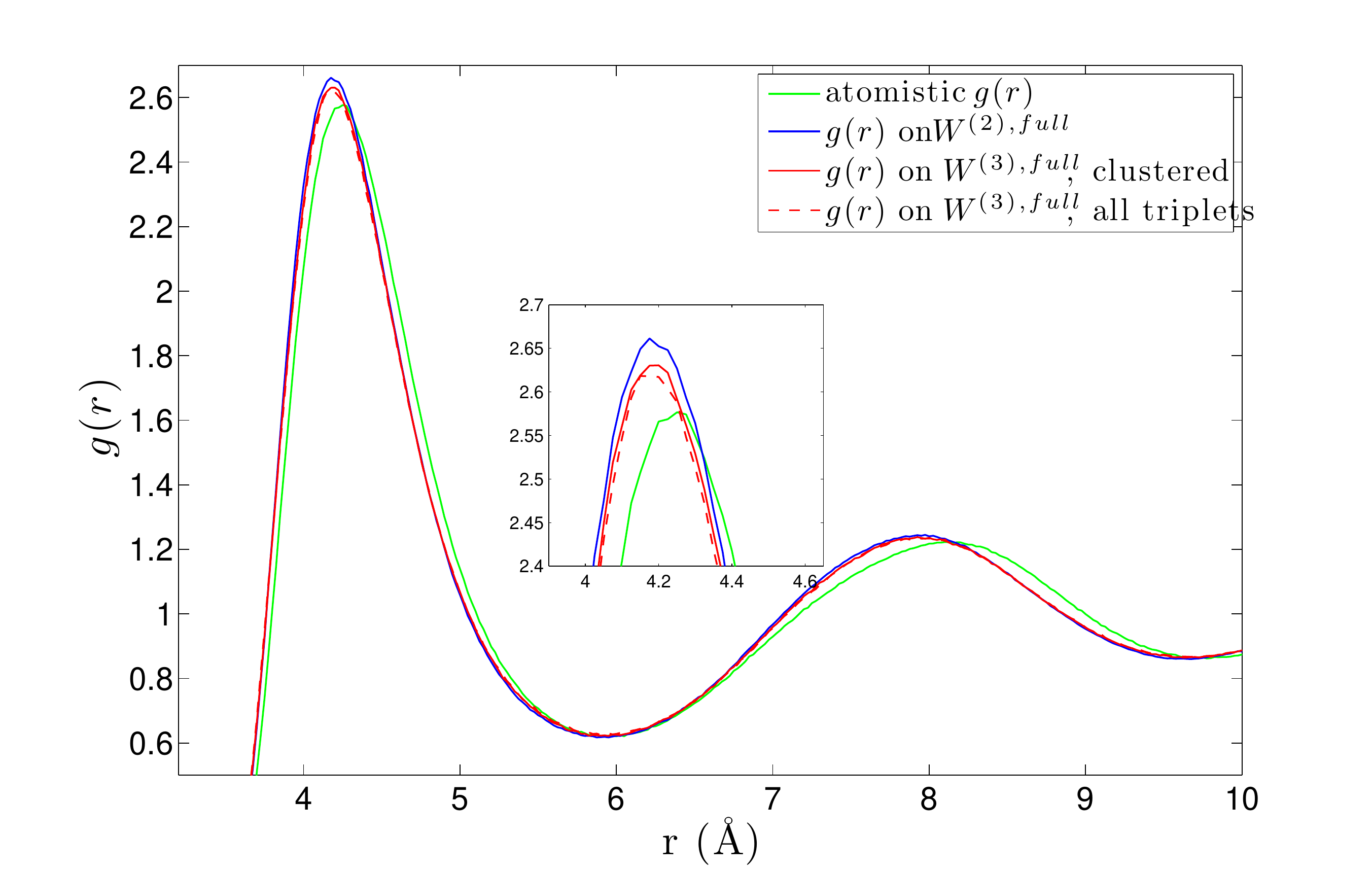}}
		\caption{RDF from atomistic and CG using pair, $W^{(2),\text{full}}$, and three-body, $W^{(3),\text{full}}$, potential for $CH_{4}$ ($T=80K$). Three dimensional cubic polynomial was used for the fitting. "Clustered" means that one particular CG atom belongs to one triplet in comparison to all possible triplets. }
	\label{fig:fig11}
\end{figure}

Next we examine the effect of the 3-body term on the CG model by performing bulk CG stochastic dynamics simulations using the new CG model with the 3-body terms described above.
Results about the pair distribution function, $g(r)$, for bulk (liquid) methane at $T=80K$ are shown in in Figure~\ref{fig:fig11}.
In this graph data from the atomistic MD runs (projected in the CG description), the CG model involving only pair CG potentials, and the new CG model that also involves 3-body terms are shown.
First, it is clear that $g(r)$ data derived from the CG model that involves only pair CG potentials show clear deviations, compared to the reference all-atom data. Note, that these differences are slightly larger than the ones discussed before (see Figure~~\ref{fig:fig7}), in which data at a higher temperature are presented.
Second, the incorporation of the three-body terms in the effective CG potential slightly improves the prediction of the $g(r)$, mainly in the first maximum regime.

\section{Discussion and conclusions}\label{sec:discussion}

In recent years we have experienced an enormous increase of computational power due to both hardware improvements and clever CPU-architecture.
However, atomistic simulations of large complex molecular systems are still out of reach in particular when long computational times are desirable.
A generic strategy in order to improve efficiency of the computational methods is to reduce  the dimensionality (degrees of freedom) by considering systematic coarse-grained models.
There have been many suggestions on how to compute the relevant CG effective interactions in such models; a main issue here is that even if in the microscopic (atomistic) level there are only pair interactions, after coarse-graining a multi-body effective potential (many-body PMF) is derived, which for realistic molecular complex systems cannot be calculated. 
Therefore, a common trend has been to approximate them by an ``effective'' pair potential by comparing the pair correlation function $g(r)$. This seems reasonable since given the correlation function one can solve the ``inverse problem'' \cite{KLS} and find an interaction to which it corresponds. But, this is an uncontrolled approximation without thermodynamic consistency.

Instead, here we suggest to explicitly compute the constrained configuration  integral over all atomistic configurations that correspond to a given coarse-grained state and from that suggest approximations with a quantifiable error.
This is similar to the virial expansion where one needs to integrate over all positions of particles that correspond to a fixed density and it is based on the recent development of establishing the cluster expansion in the canonical ensemble.~\cite{PT12}; see also Ref.~\cite{KT,McQ} for the corresponding (in the canonical ensemble) expansions for the correlation functions and the Ornstein-Zernike equation.
The main drawback that limits the applicability of these expansions is that they are rigorously valid only in the gas phase. To extend them to the liquid state is an outstanding problem and even several successful closures like the Percus-Yevick are not rigorously justified. Therefore, there is need of further developing these methods and relate them to computational strategies. 

In this paper we extend the above methods by presenting an approach based on cluster expansion techniques and numerical computations of isolated molecules. 
As a first test we presented a detailed investigation of the proposed methodology to derive CG potentials for methane and ethane molecular systems. Each CG variable corresponds to the center-of-mass for each molecule.
Below, we summarize our main findings:

(a) The hierarchy of the cluster expansion formalism allowed us to systematically define the CG effective interaction as a sum of pair, triplets, etc. interactions. Then, CG effective potentials can be computed as they arise from the cluster expansion.

(b) The two-body coarse-grained potentials can be efficiently computed via the cluster expansion giving comparable results with the existing methods, such as the conditional reversible work. In addition we present a more efficient direct geometric computation of the constrained partition function.

(c) The obtained pair CG potentials were used to model the corresponding liquid systems and the derived $g(r)$ data were compared against the all-atom ones. Clear differences between methane and ethane systems were observed; For the (almost spherical) methane, pair CG potentials seems to be a very good approximation, whereas much larger differences between CG and atomistic distribution functions were observed for ethane.  

(d) We further investigated different temperature and density regimes, and in particular cases where the two-body approximations are not good enough compared to the atomistic simulations. In the latter case, we considered the next term in the cluster expansion, namely the three-body effective potentials and we found that they give a small improvement over the pair ones.

Overall, we conjecture that the cluster expansion formalism can be used in order to provide accurate effective pair and three-body CG potentials at high $T$ and low $\rho$ regimes. 
In order to get significantly better results in the {\it liquid} regime one needs to consider even higher order terms, which are in general more expensive to be computed and more difficult to be treated. A more detailed analysis of the higher-order terms will be a part of a future work.~\cite{THT2017}
Finally, another future goal is to extend this investigation in larger molecules (e.g. polymeric chains) that involve intra-molecular CG effective interactions as well, and to systems with long range (e.g. Coulombic) interactions.




\begin{acknowledgments}
We acknowledge support by the European Union (ESF) and Greek national funds through the Operational Program “Education and Lifelong Learning” of the NSRF-Research Funding Programs: THALES and ARISTEIA II. 
\end{acknowledgments}

\bibliographystyle{unsrt}

\bibliography{references}

\begin{thebibliography}{10}

\bibitem{FrenkelSmitBook}
D.~Frenkel and B.~Smit.
\newblock {\em Understanding Molecular Simulation: From Algorithms to
  Applications}.
\newblock Academic Press, second edition, 2001.

\bibitem{AllenTildesleyBook}
M.~P. Allen and D.~J. Tildesley.
\newblock {\em Computer Simulation of Liquids}.
\newblock Oxford University Press, 1987.

\bibitem{Harmandaris2003a}
V.~A. Harmandaris, V.~G. Mavrantzas, D.~Theodorou, M.~Kröger, J.~Ramírez,
  H.~C. \"Ottinger, and D.~Vlassopoulos.
\newblock Dynamic crossover from rouse to entangled polymer melt regime:
  Signals from long, detailed atomistic molecular dynamics simulations,
  supported by rheological experiments.
\newblock {\em Macromolecules}, 36:1376--1387, 2003.

\bibitem{TheodorouBook}
M.~Kotelyanskii and D.~N. Theodorou.
\newblock {\em Simulation Methods for Polymers}.
\newblock Taylor \& Francis, 2004.

\bibitem{IzVoth2005a}
S.~Izvekov and G.~A. Voth.
\newblock Multiscale coarse graining of liquid-state systems.
\newblock {\em J. Chem. Phys.}, 123(13):134105, 2005.

\bibitem{tsop1}
W.~Tsch{\"o}p, K.~Kremer, O.~Hahn, J.~Batoulis, and T.~B{\"u}rger.
\newblock Simulation of polymer melts. {I}. coarse-graining procedure for
  polycarbonates.
\newblock {\em Acta Polym.}, 49:61, 1998.

\bibitem{MulPlat2002}
F.~M\"uller-Plathe.
\newblock Coarse-graining in polymer simulation: From the atomistic to the
  mesoscopic scale and back.
\newblock {\em ChemPhysChem}, 3(9):754--769, 2002.

\bibitem{Shell2008}
M.~S. Shell.
\newblock The relative entropy is fundamental to multiscale and inverse
  thermodynamic problems.
\newblock {\em J. Chem. Phys.}, 129(14):--, 2008.

\bibitem{briels}
W.~J. Briels and R.~L.~C. Akkermans.
\newblock {Coarse-grained interactions in polymer melts: a variational
  approach}.
\newblock {\em J. Chem. Phys.}, 115:6210, 2001.

\bibitem{Harmandaris2006a}
V.~A. Harmandaris, N.~P. Adhikari, N.~F.~A. van~der Vegt, and K.~Kremer.
\newblock Hierarchical modeling of polystyrene: From atomistic to
  coarse-grained simulations.
\newblock {\em Macromolecules}, 39:6708, 2006.

\bibitem{Harmandaris2009a}
V.~A. Harmandaris and K.~Kremer.
\newblock Dynamics of polystyrene melts through hierarchical multiscale
  simulations.
\newblock {\em Macromolecules}, 42:791, 2009.

\bibitem{Harmandaris2009b}
V.~A. Harmandaris and K.~Kremer.
\newblock Predicting polymer dynamics at multiple length and time scales.
\newblock {\em Soft Matter}, 5:3920, 2009.

\bibitem{Johnston2013}
K.~Johnston and V.~Harmandaris.
\newblock Hierarchical simulations of hybrid polymer– solid materials.
\newblock {\em Soft Matter}, 9:6696--6710, 2013.

\bibitem{Voth2008a}
W.~G. Noid, J.-W. Chu, G.~S. Ayton, V.~Krishna, S.~Izvekov, G.~A. Voth, A.~Das,
  and H.~C. Andersen.
\newblock The multiscale coarse-graining method. i. a rigorous bridge between
  atomistic and coarse-grained models.
\newblock {\em J. Chem. Phys.}, 128(24):4114, 2008.

\bibitem{Voth2010}
L.~Lu, S.~Izvekov, A.~Das, H.~C. Andersen, and G.~A. Voth.
\newblock Efficient, regularized, and scalable algorithms for multiscale
  coarse-graining.
\newblock {\em Journal of Chemical Theory and Computation}, 6(3):954--965,
  2010.

\bibitem{Noid2011}
J.~F. Rudzinski and W.~G. Noid.
\newblock Coarse-graining entropy, forces, and structures.
\newblock {\em J. Chem. Phys.}, 135(21):214101, 2011.

\bibitem{Noid2013}
W.~G. Noid.
\newblock Perspective: Coarse-grained models for biomolecular systems.
\newblock {\em J. Chem. Phys.}, 139(9):090901, 2013.

\bibitem{Shell2009}
A.~Chaimovich and M.~S. Shell.
\newblock Anomalous waterlike behavior in spherically-symmetric water models
  optimized with the relative entropy.
\newblock {\em Phys. Chem. Chem. Phys.}, 11:1901--1915, 2009.

\bibitem{Zabaras2013}
I.~Bilionis and N.~Zabaras.
\newblock A stochastic optimization approach to coarse-graining using a
  relative-entropy framework.
\newblock {\em J. Chem. Phys.}, 138(4):--, 2013.

\bibitem{DiffusionMapKevrekidis2008}
R.~R. Coifman, I.~G. Kevrekidis, S.~Lafon, M.~Maggioni, and B.~Nadler.
\newblock Diffusion maps, reduction coordinates, and low dimensional
  representation of stochastic systems.
\newblock {\em Multiscale Modeling \& Simulation}, 7(2):842--864, 2008.

\bibitem{Soper1996}
A.~K. Soper.
\newblock Empirical potential monte carlo simulation of fluid structure.
\newblock {\em Chemical Physics}, 202(203):295 -- 306, 1996.

\bibitem{LyubLaa2004}
A.~P. Lyubartsev and A.~Laaksonen.
\newblock {On the Reduction of Molecular Degrees of Freedom in Computer
  Simulations}.
\newblock In M.~Karttunen, A.~Lukkarinen, and I.~Vattulainen, editors, {\em
  Novel Methods in Soft Matter Simulations}, volume 640 of {\em Lecture Notes
  in Physics, Berlin Springer Verlag}, pages 219--244, 2004.

\bibitem{VagelisReview2014}
V.~A. Harmandaris.
\newblock Quantitative study of equilibrium and non-equilibrium polymer
  dynamics through systematic hierarchical coarse-graining simulations.
\newblock {\em Korea-Australia Rheology Journal}, 26(1):15--28, 2014.

\bibitem{EspanolZuniga2011}
P.~Espanol and I.~Zuniga.
\newblock Obtaining fully dynamic coarse-grained models from md.
\newblock {\em Phys. Chem. Chem. Phys.}, 13:10538--10545, 2011.

\bibitem{BrielsPadding}
J.~T. Padding and W.~J. Briels.
\newblock Uncrossability constraints in mesoscopic polymer melt simulations:
  Non-rouse behavior of c120h242.
\newblock {\em J. Chem. Phys.}, 115(6):2846--2859, 2001.

\bibitem{Nico2014}
G.~Deichmann, V.~Marcon, and N.~F.~A. van~der Vegt.
\newblock Bottom-up derivation of conservative and dissipative interactions for
  coarse-grained molecular liquids with the conditional reversible work method.
\newblock {\em J. Chem. Phys.}, 141:224109, 2014.

\bibitem{Fritz2009}
D.~Fritz, V.~Harmandaris, K.~Kremer, and N.~van~der Vegt.
\newblock Coarse-grained polymer melts based on isolated atomistic chains:
  Simulation of polystyrene of different tacticities.
\newblock {\em Macromolecules}, 42:7579--7588, 2009.

\bibitem{Nico-Review2013}
Emiliano Brini, Elena~A. Algaer, Pritam Ganguly, Chunli Li, Francisco
  Rodr´ıguezRopero, and Nico F.~A. van~der Vegt.
\newblock Systematic coarse-graining methods for soft matter simulations–a
  review.
\newblock {\em Soft Matter}, 9:2108–2119, 2013.

\bibitem{Reith2003}
D.~Reith, M.~P{\"u}tz, and F.~M{\"u}ller-Plathe.
\newblock Deriving effective mesoscale potentials from atomistic simulations.
\newblock {\em Journal of computational chemistry}, 24(13):1624--1636, 2003.

\bibitem{Lyubartsev1995}
A.~P. Lyubartsev and A.~Laaksonen.
\newblock {Calculation of effective interaction potentials from radial
  distribution functions: A reverse Monte Carlo approach}.
\newblock {\em Phys. Rev. E}, 52:3730--3737, 1995.

\bibitem{IzVoth2005}
S.~Izvekov and G.~A. Voth.
\newblock Effective force field for liquid hydrogen fluoride from ab initio
  molecular dynamics simulation using the force-matching method.
\newblock {\em The Journal of Physical Chemistry. B}, 109(14):6573--6586, 04
  2005.

\bibitem{Voth2008b}
W.~G. Noid, P.~Liu, Y.~Wang, J.~Chu, G.~S. Ayton, S.~Izvekov, H.~C. Andersen,
  and G.~A. Voth.
\newblock The multiscale coarse-graining method. ii. numerical implementation
  for coarse-grained molecular models.
\newblock {\em J. Chem. Phys.}, 128(24):244115, 2008.

\bibitem{Voth2013}
L.~Lu, J.~F. Dama, and G.~A. Voth.
\newblock Fitting coarse-grained distribution functions through an iterative
  force-matching method.
\newblock {\em J. Chem. Phys.}, 139:121906, 2013.

\bibitem{KP2013}
M.~A. Katsoulakis and P.~Plechac.
\newblock Information-theoretic tools for parametrized coarse-graining of
  non-equilibrium extended systems.
\newblock {\em J. Chem. Phys.}, 139:4852--4863, 2013.

\bibitem{Shell2011}
A.~Chaimovich and M.~S. Shell.
\newblock Coarse-graining errors and numerical optimization using a relative
  entropy framework.
\newblock {\em J. Chem. Phys.}, 134(9):094112, 2011.

\bibitem{Chu2009}
H.~M. Cho and J.~W. Chu.
\newblock Inversion of radial distribution functions to pair forces by solving
  the yvon–born–green equation iteratively.
\newblock {\em J. Chem. Phys.}, 131:134107, 2009.

\bibitem{Noid2007}
W.~G. Noid, J.~Chu, G.~S. Ayton, and G.~A. Voth.
\newblock Multiscale coarse graining and structural correlations: Connections
  to liquid-state theory.
\newblock {\em J. Phys. Chem. B}, 111:4116–4127, 2007.

\bibitem{Noid2009}
J.~W. Mullinax and W.~G. Noid.
\newblock Generalized yvon–born–green theory for molecular systems.
\newblock {\em Phys. Rev. Lett.}, 103:198104, 2009.

\bibitem{Noid2010}
J.~W. Mullinax and W.~G. Noid.
\newblock Generalized yvon–born–green theory for determining coarse-grained
  interaction potentials.
\newblock {\em J. Phys. Chem. C}, 114:5661–5674, 2010.

\bibitem{Guenza2014}
J.~McCarty, A.~J. Clark, J.~Copperman, and M.~G. Guenza.
\newblock An analytical coarse-graining method which preserves the free energy,
  structural correlations, and thermodynamic state of polymer melts from the
  atomistic to the mesoscale.
\newblock {\em J. Chem. Phys.}, 140(20):--, 2014.

\bibitem{Karniadakis2015}
Z.~Li, X.~Bian, X.~Li, and G.~E. Karniadakis.
\newblock Incorporation of memory effects in coarse-grained modeling via the
  mori-zwanzig formalism.
\newblock {\em J. Chem. Phys.}, 143(-):243128, 2015.

\bibitem{Guenza2013}
A.~J. Clark, J.~McCarty, and M.~G. Guenza.
\newblock Effective potentials for representing polymers in melts as chains of
  interacting soft particles.
\newblock {\em J. Chem. Phys.}, 139:124906, 2013.

\bibitem{MM}
J.~E. Mayer and M.~G. Mayer.
\newblock {\em Statistical Mechanics}.
\newblock John Wiley and Sons, 1940.

\bibitem{MoHi61}
T.~Morita and K.~Hiroike.
\newblock {The statistical mechanics of condensing systems. III}.
\newblock {\em Prog. Theor. Phys.}, 25:537, 1961.

\bibitem{Stell64}
G.~Stell.
\newblock {\em Cluster expansion for classical systems in equilibrium in H.
  Frisch and J. Lebowitz, ed., Classical Fluids}.
\newblock New York: Benjamin, 1964.

\bibitem{PT12}
E.~Pulvirenti and D.~Tsagkarogiannis.
\newblock {Cluster expansion in the canonical ensemble}.
\newblock {\em Comm. Math. Phys.}, 316(2):289--306, 2012.

\bibitem{KPRT07}
M.~Katsoulakis, P.~Plech\'a\v{c}, L.~Rey-Bellet, and D.~Tsagkarogiannis.
\newblock {Coarse-graining schemes and a posteriori error estimates for
  stochastic lattice systems}.
\newblock {\em ESAIM: Math. Model. and Num. Analysis}, 41(3):627--660, 2007.

\bibitem{KPRT14}
M.~Katsoulakis, P.~Plech\'a\v{c}, L.~Rey-Bellet, and D.~Tsagkarogiannis.
\newblock { Coarse-graining schemes for stochastic lattice systems with short
  and long range interactions}.
\newblock {\em Mathematics of Computation (AMS)}, 83(-):1757--1793., 2014.

\bibitem{KPRT08}
M.~Katsoulakis, P.~Plech\'a\v{c}, L.~Rey-Bellet, and D.~Tsagkarogiannis.
\newblock { Mathematical strategies and error quantification in coarse graining
  of extended systems}.
\newblock {\em J. Non-Newtonian Fluid Mech.}, 152(-):101--112, 2008.

\bibitem{TT10}
J.~Trashorras and D.~Tsagkarogiannis.
\newblock {Reconstruction schemes for coarse-grained stochastic lattice
  systems}.
\newblock {\em SIAM J. Numer. Anal.}, 48(5):1647--1677, 2010.

\bibitem{KM_P}
K.~Kremer and F.~M{\"u}ller-Plathe.
\newblock {Multiscale problems in polymer science: simulation approaches}.
\newblock {\em MRS Bull.}, -:205, 2001.

\bibitem{McCoy1998}
J.~D. McCoy and J.~G. Curro.
\newblock Mapping of explicit atom onto united atom potentials.
\newblock {\em Macromolecules}, 31:9352--9368, 1998.

\bibitem{THT2017}
A.~Tsourtis, V.~Harmandaris, and D.~Tsagkarogiannis.
\newblock Effective coarse-grained interactions: The role of three-body terms
  through cluster expansions.
\newblock {\em under preparation}.

\bibitem{Kirkwood1935}
J.~G. Kirkwood.
\newblock Statistical mechanics of fluid mixtures.
\newblock {\em J. Chem. Phys.}, 3(5):300--313, 1935.

\bibitem{McQ}
D.~A. McQuarrie.
\newblock {\em Statistical Mechanics}.
\newblock University Science Books, 2000.

\bibitem{Guenza}
J.~McCarty, A.~J. Clark, J.~Copperman, and M.~G. Guenza.
\newblock An analytical coarse-graining method which preserves the free energy,
  structural correlations, and thermodynamic state of polymer melts from the
  atomistic to the mesoscale.
\newblock {\em J. Chem. Phys.}, 140(20):--, 2014.

\bibitem{KCTKPH2016}
E.~Kalligiannaki, A.~Chazirakis, A.~Tsourtis, M.~Katsoulakis,
  P.~Plech{\'a}\v{c}, and V.~Harmandaris.
\newblock Parametrizing coarse grained models for molecular systems at
  equilibrium.
\newblock {\em European Physical Journal, Special Topics}, 225:1347–1372,
  2016.

\bibitem{Louis02}
A.~A. Louis.
\newblock {Beware of density dependent pair potentials}.
\newblock {\em J.Phys.: Condens. Matter}, 14:9187–9206, 2002.

\bibitem{Louis01}
P.~G. Bolhuis, A.~A. Louis, and J.~P. Hansen.
\newblock Many-body interactions and correlations in coarse-grained
  descriptions of polymer solutions.
\newblock {\em Phys. Rev. E}, 64:021801, 2001.

\bibitem{PT14}
E.~Pulvirenti and D.~Tsagkarogiannis.
\newblock {Finite volume corrections and decay of correlations in the canonical
  ensemble}.
\newblock {\em J. Stat. Phys.}, 159(5):1017--1039, 2014.

\bibitem{KT}
T.~Kuna and D.~Tsagkarogiannis.
\newblock {Convergence of density expansions of correlation functions and the
  Ornstein-Zernike equation}.
\newblock {\em preprint arXiv:1611.01716}, 2016.

\bibitem{Dreiding1990}
S.~Mayo, B.~Olafson, and W.~Goddard.
\newblock Dreiding: a generic force field for molecular simulations.
\newblock {\em J. Phys. Chem.}, 94(26):8897--8909, 1990.

\bibitem{Wick_Siepmann_Trappe_ETH}
C.~D. Wick, M.~G. Martin, and J.~I. Siepmann.
\newblock Transferable potentials for phase equilibria. 4. united-atom
  description of linear and branched alkenes and alkylbenzenes.
\newblock {\em J. Phys. Chem. B}, 104:8008--8016, 2000.

\bibitem{lelievre2010}
T.~Leli{\`e}vre, M.~Rousset, and G.~Stoltz.
\newblock {\em Free Energy Computations: A Mathematical Perspective}.
\newblock Imperial College Press, 2010.

\bibitem{KLS}
T.~Kuna, J.~Lebowitz, and E.~Speer.
\newblock {Realizability of point processes}.
\newblock {\em J. Stat. Phys.}, 129:417--439, 2007.

\end{thebibliography}

\end{document}